\documentclass[12pt]{article}
\usepackage{fancybox,amssymb, graphicx,graphics,color,epsfig,subfigure,latexsym,tabularx,times,rotating,amsmath,ctable}
\usepackage{subfigure}
\usepackage{lscape}

\def\sign{\mathop{\rm sign}\nolimits}

\def\argmax{\mathop{\rm argmax}\nolimits}
\def\argmin{\mathop{\rm argmin}\nolimits}

\begin{document}

\begin{center}

$\;\;$
\hrule
\hrule
$\;\;$

\vspace{1cm}
{\bf\large{Reconstructing DNA Copy Number by Joint Segmentation of Multiple Sequences}} 
\vspace{0.5cm}

Zhongyang Zhang, Kenneth Lange, and Chiara Sabatti
\vspace{1cm}  


\hrule
\hrule

\vspace{1cm}


March 2012

\vspace{1cm}

\end{center}

\begin{abstract}
Variations in DNA copy number carries information on the modalities of genome evolution and misregulation of DNA replication in cancer cells; their study can be helpful to localize tumor suppressor genes, distinguish different populations of cancerous cell, as well identify genomic variations responsible for disease phenotypes. A number of different high throughput technologies can be used to identify copy number variable sites, and the literature documents multiple effective algorithms. We focus here on the specific problem of detecting regions where variation in copy number is relatively common in the sample at hand: this encompasses the cases of copy number polymorphisms, related samples, technical replicates, and cancerous sub-populations from the same individual. We present an algorithm based on regularization approaches with significant computational advantages and competitive accuracy. We illustrate its applicability with simulated and real data sets.
\end{abstract}

\section{Introduction}

Duplication and deletion of genomic materials are common in cancer cells and known to play a role in the establishment of the tumor status \cite{newton2000}. As our ability to survey the fine scale of the human genome has increased, it has become apparent that normal cells can also harbor a number of variations in copy number \cite{Iafrate04, sebat04}. The last few years have witnessed a steady increase in our knowledge of size and frequency of these variants \cite{sanger_cnv, HapMapCNV, seq_cnv, redon06} and their implications in complex diseases \cite{Nature_Autism_CNV, stefansson08_schizo}.

At the same time, statistical methods and algorithms have been developed to better harness the information available. At the cost of oversimplification, two different approaches have become particularly popular: one is based on the hidden Markov model (HMM) machinery and explicitly aims to reconstruct the unobservable discrete DNA copy number; the other, which we will generically call ``segmentation", aims at identifying portions of the genome that have constant copy number, without specifically reconstructing it. The HMM approach takes advantage of the implicitly discrete nature of the copy number process (both when a finite number of states is assumed and when, as in some implementations, less parametric approaches are adopted); furthermore, by careful modeling of the emission probabilities, one can fully utilize the information derived from the experimental results. In the case of genotyping arrays, for example, both the quantification of total DNA amount and relative allelic abundance as well as prior information (for example, minor allele frequencies) can be considered. No a-priori knowledge of the number of copy number states is required the segmentation approach---an advantage in the study of cancer where polyploids and contamination with normal tissues result in a wide range of fractional copy numbers.  Possibly for the reasons outlined, HMMs are the methods of choice in the analysis of normal samples \cite{QuantiSNP, ingo, weisun, PennCNV, holmes2011}, while segmentation methods are the standard in cancer studies \cite{CBS, Nancy_Biometrika10}. A limitation of segmentation methods is that they rely on the data in which the variation in copy number is reflected in the differences in means of the segments---which make them applicable directly to a substantial portion of the data derived from recent technologies, but not to relative allelic abundance (see the modification suggested in \cite{Staaf08} and following description for an exception).

While a number of successful approaches have been derived along the lines described above, there is still a paucity of methodology for the joint analysis of multiple sequences.  It is clear that if multiple subjects share the same variation in copy number, there exists the potential to increase power by joint analysis. Wang et al. (2009) \cite{us} presented a methodology that extended \cite{newton2000} to reconstruct the location of tumor suppressor genes from the identification of regions lost in a larger number of samples; the initial steps of the Birdsuite algorithm rely on the identification of suspect signals in the context of multiple samples; PennCNV \cite{PennCNV} includes an option of joint analysis of trios; methodology to process multiple samples with the context of change point analysis has been developed in  \cite{Siegmund_AOAS11, Nancy_Bioinformatics10, Nancy_Biometrika10}; Efron and Zhang (2011) \cite{efron_nancy_fdr} consider FDR analysis of independent samples to identify copy number polymorphysms (CNPs); and Nowak et al. (2011) \cite{FLLat} use a latent feature model to capture, in joint analysis of array-CGH data from multiple tumor samples,  shared copy number profiles, on each of which a fused-lasso penalty is enforced for sparsity.
In the present work we consider a setting similar to \cite{Nancy_Biometrika10} in that we want joint analysis to inform the segmentation of multiple samples. Our main focus is the analysis of genotyping array data, but the methodology we develop is applicable to a variety of platforms. By adopting a flexible framework we are able, for example, to define a segmentation algorithm that uses all information from Illumina genotyping data. As in \cite{Siegmund_AOAS11}, we are interested in the situation when not all the samples under consideration carry a copy number variant (CNV): we rather want to enforce a certain sparsity in the vector that identifies which samples carry a given variant. We tackle this problem using a penalized estimation approach, originally proposed in this context by \cite{cghFLasso}, on which we have developed an algorithmic implementation before \cite{FL_DPI}. Appreciable results are achieved in terms of speed, accuracy and flexibility. 
In concluding this introduction, we would like to make an important qualification: the focus of our contribution is on segmentation methods, knowing that this is only one of the steps necessary for an effective recovery of CNVs. In particular, normalization and transformation of the signal from experimental sources are crucial and can have a very substantial impact on final results: we refer the reader to \cite{Bengtsson2010, Bengtsson2009,  Bengtsson2007, gcadj, crlmm_Scharpf2011, Scharpf2011}, for example. Furthermore, calling procedures that further classify results of segmentation while possibly controlling global error measures \cite{efron_nancy_fdr} are also needed. Indeed, in the data analysis included in this paper we need to resort to both these additional steps and we will describe briefly the fairly standard choices we are making.

The rest of the paper is organized as follows: Section \ref{methods} motivates the need for joint analysis of multiple signals and presents the penalized estimation framework. Section \ref{alg} describes how the model can be used for data analysis by (a) outlining an efficient estimation algorithm, (b) generalizing it to the case of uncoordinated data, and (c) describing the choice of the penalization parameters. Section \ref{results} illustrates our results on two simulated data sets (descriptive of normal and tumor samples) and two real data sets: in one case multiple platforms are used to analyze the same sample and in the other case samples from related individuals benefit from joint analysis. 

\section{Multiple sequence segmentation} \label{methods}
The goal of the present paper is to develop a flexible methodology for joint segmentation of multiple sequences that are presumed to carry related information on CNVs. We start by illustrating a series of contexts where the joint analysis appears to be useful.

\subsection{Motivation}

\subsubsection{Genotyping arrays and CNV detection}
Genotyping arrays have been used on hundreds of thousands of subjects and the data collected through them provides an extraordinary resource for CNV detection and the study of their frequencies in multiple populations. Typically, the raw intensity data representing hybridization strength is processed to obtain two signals: a quantification of total DNA amount (from now on log R Ratio LRR, following Illumina terminology) and a relative abundance of the two queried alleles (from now on B allele frequency, BAF).  Both these signals contain information on CNV and one of the strengths of HMM models has been that they can easily process them jointly. Segmentation models like CBS have traditionally relied only on LRR. While this is a reasonable choice, it can lead to substantial loss of information, particularly in tumor cells, where poliploidity and contamination make information in LRR hard to decipher. To exploit BAF in the context of a segmentation method, a signal transformation has been suggested \cite{Staaf08}: mirrowed BAF (mBAF) relies on exchangeability of the two alleles and the low information content of homozygous SNPs. The resulting mBAF is defined on a coarser grid than the original BAF, but is characterized by changing means in presence of CNV. While \cite{Staaf08}  shows that its analysis alone can be advantageous and more powerful than segmentation of LRR in some contexts, clearly a joint analysis of LRR and mBAF should be preferable to an arbitrary selection of one or the other signal. 


\subsubsection{Multiple platforms}
LRR and BAF are just one example of the multiple signals that one can have available for the same sample. Often, as research progresses, the samples are assessed with a variety of technologies. For example, a number of subjects that have been genotyped at high resolution are now being resequenced. Whenever the technology adopted generates a signal that contains some information on copy number, there is an incentive to analyze the available signals jointly.

\subsubsection{Tumor samples from the same patient obtained at different sites or different progression stages} \label{eg_tumor}
In an effort to identify mutations that are driving a specific tumor, as well as study its response to treatment, researchers might want to study CNVs in cells obtained at different tumor sites or at different time points \cite{tumor_SIM10}. Copy number is highly dynamic in cancer cells, so that it is to be expected that some differences be detected over time or across sites. In contrast, the presence of the same CNVs across these samples, can be taken as an indication that the tumors share the same origin: therefore a comparative analysis of CNV can be used to distinguish resurgence of the same cancer from insurgence of a new one, or to identify specific cancer cell populations. Given that the tissue extracted always consists of a mixture of normal and cancer cells, which are in turn a mixture of different populations, joint analysis of the signals from the varied materials is much more likely to lead to the identification of common CNVs, when these exist.

\subsubsection{Related subjects} \label{eg_related}
 
Family data is crucial in genetic investigations and hence it is common to analyze related subjects. When studying individuals from the same pedigree, it is reasonable to assume that some CNVs might be segregating in multiple people: joint analysis would reduce Mendelian errors and increase power of detection.

\subsection{A model for joint analysis of multiple signals}

Assume we have observed $M$ signals, each measured at $N$ locations, corresponding to ordered physical positions along the genome, with $y_{ij}$ being the observed value of sequence $i$ at location $j$.
The copy number process can be modeled as 
\begin{equation} \label{piece_const}
y_{ij} = \beta_{ij} + \epsilon_{ij},
\end{equation}
where $\epsilon_{ij}$ represent noise, and the mean values $\beta_{ij}$ are piece-wise constant: there exists a linearly ordered partition $\{R_1^{(i)},R_2^{(i)},\ldots,R_{K_i}^{(i)}\}$ of the location index $\{1,2,\ldots,N\}$ such that $\beta_{is} = \cdots = \beta_{it} = \mu_k^{(i)}$ for $s,\ldots,t \in R_k^{(i)}$ and $1\leq k\leq K_i$. In other words, most of the increments $|\beta_{ij}-\beta_{i,j-1} |$ are assumed to be zero. When two sequences $k$ and $l$ share a CNV with the same boundaries at location $j$, both $|\beta_{kj}-\beta_{k,j-1} |$ and $|\beta_{lj}-\beta_{l,j-1} |$ 
will be different from zero in correspondence of the change point.
Modulo an appropriate signal normalization, $\beta_{ij}=0$ can be interpreted as corresponding to the appropriate normal copy number equal to 2.
We propose to reconstruct the mean values $\boldsymbol{\beta}$ by minimizing the following function, called hereafter generalized fused lasso (GFL):

\begin{equation} \label{model}
f(\boldsymbol{\beta}) = \frac{1}{2}\sum_{i=1}^M\sum_{j=1}^N (y_{ij}-\beta_{ij})^2 + \lambda_{1}\sum_{i=1}^M  \sum_{j=1}^N |\beta_{ij}| + \lambda_{2}\sum_{i=1}^M\sum_{j=2}^N |\beta_{ij} - \beta_{i,j-1}| +  \lambda_{3 }\sum_{j=2}^N \left[\sum_{i=1}^M (\beta_{ij}-\beta_{i,j-1})^2\right]^{\frac{1}{2}},
\end{equation}
which includes a goodness-of-fit term and three penalties, whose roles we will explain one at the time. 
The $\ell_1$ penalty $\sum_{i=1}^M \sum_{j=1}^N |\beta_{ij}| $ enforces sparsity within $\boldsymbol{\beta}$, in favor of values $\beta_{ij}=0$, corresponding to the normal copy number. The total variation penalty $\sum_{j=2}^N |\beta_{ij} - \beta_{i,j-1}|$ minimizes the number of jumps in the piece-wise constant means of each sequence and was introduced by \cite{cghFLasso} in the context of CNV reconstruction from array-CGH data. Finally, the Euclidean penalty on the column vector of jumps $\sqrt{\sum_{i=1}^M (\beta_{ij}-\beta_{i, j-1})^2}$ is a form of the group penalty introduced by \cite{GroupLasso} and favors common jumps across sequences. As clearly explained in \cite{hua10}, ``the local penalty around 0 for each member of a group relaxes as soon as the $|\beta_{ij}-\beta_{i,j-1}|$ for one  member $i$ of the group moves off 0." Bleakley and Vert (2011)  \cite{GroupFL_new} also suggested the use of this group-fused-lasso penalty to reconstruct CNV.  We here consider the use of both the total variation and the Euclidean penalty on the jumps to achieve the equivalent effect of the sparse group lasso, which, as pointed out in \cite{sparse_GL}, favors CNV detection in multiple samples,  allowing for sparsity in the vector indicating which subjects are carriers of the variant. This property is important in situations as presented in Section \ref{eg_tumor} and \ref{eg_related}, where one does not want to assume that all the $M$ sequences carry the same CNV.

The incorporation of the latter two penalties can also be naturally interpreted in view of image denoising. To restore an image disturbed by random noise while preserving sharp edges of items in the image, a 2-D total variation penalty $\lambda\sum_{i=1}^M\sum_{j=2}^N |\beta_{ij}-\beta_{i,j-1}| + \rho \sum_{j=1}^N\sum_{i=2}^M |\beta_{ij}-\beta_{i-1,j}|$ is proposed in a regularized least-square optimization \cite{TV_denoise}, where $\beta_{ij}$ is the true underlying intensity of pixel $(i, j)$. In CNV detection problems, signals from multiple sequences can be aligned up in shape of an image, except that pixels in each sequence are linearly ordered while sequences as a group have no certain order a priori; thus one of the two total variation penalties is replaced by the group penalty on the column vector of jumps.

Using matrix notation, and allowing the tuning parameter $\lambda_1$, $\lambda_2$ and $\lambda_3$ to be sequence specific, we can reformulate the objective function as follows. Let $\mathbf{Y}=(y_{ij})_{M\times N}$ and $\boldsymbol{\beta}=(\beta_{ij})_{M\times N}$. Let $\boldsymbol{\beta}_i$ be the $i$th row of $\boldsymbol{\beta}$ and $\boldsymbol{\beta}_{(j)}$ the $j$th column of $\boldsymbol{\beta}$. Also, let $\boldsymbol{\lambda}_3 = (\lambda_{3,i})_{M \times 1}$. Then we have
\begin{eqnarray}
f(\boldsymbol{\beta}) &=& \frac{1}{2}||\mathbf{Y} - \boldsymbol{\beta}||_F^2 + \sum_{i=1}^M \lambda_{1,i}||\boldsymbol{\beta}_i||_{\ell_1} \nonumber \\
& & + \sum_{i=1}^M \lambda_{2,i}||\boldsymbol{\beta}_{i,2:N} - \boldsymbol{\beta}_{i,1:(N-1)}||_{\ell_1} +  \sum_{j=2}^N ||\boldsymbol{\lambda}_3 * (\boldsymbol{\beta}_{(j)} - \boldsymbol{\beta}_{(j-1)})||_{\ell_2}, \label{mmodel}
\end{eqnarray}
where $||\cdot||_F$ is the Frobenius norm of matrix, $||\cdot||_{\ell_1}$ and $||\cdot||_{\ell_2}$ are $\ell_1$ and $\ell_2$ norm of vector, $\boldsymbol{\beta}_{i,s:t}$ indicates the sub-vector with elements $\beta_{i,s},\ldots,\beta_{i,t}$ in row vector $\boldsymbol{\beta}_i$, and ``$*$" is used as entry-wise multiplication between two vectors.
Note that it would be easy to modify the tuning parameters so as to make them location specific: that is, reduce the penalty for a jump in correspondence of genomic regions known to harbor CNVs.

\section{Implementation} \label{alg}

\subsection{An MM algorithm}
While the solution to the optimization problem (\ref{mmodel}) might have interesting properties, this approach is useful only if an effective algorithm is available. The last few years have witnessed substantial advances in computational methods for $\ell_1$-regularization problems, including the use of  coordinate descent \cite{pco, wulange} and path following methods \cite{GroupFL_new,  path_hoefling, path_ryan, path_hua_ken}. The time cost of these methods in the best situation is $O(MNK)$, for $K$ knots along the solution path. It is important to note that these algorithms -- some of which are designed for more general applications -- may not be the most efficient for large scale CNV analysis for at least two reasons: on the one hand, reasonable choices of $\lambda$ might be available, making it unnecessary to solve for the entire path; on the other hand,  the number of knots $K$ can be expected to be as large as  $O(N)$, making the computational costs of path algorithms prohibitive.

With specific regard to the fused-lasso application to CNV detection, we were successful in developing algorithm with per iteration cost $O(N)$ and empirically fast convergence rate for the analysis of one sequence \cite{FL_DPI}. We apply the same principles here. We start by modifying the norms in the penalty as follows: rather than the $\ell_1$ norm  we use $||x||_{2,\epsilon} = \sqrt{x^2+\epsilon}$ for sufficiently small $\epsilon$, and, for computational stability, we also substitute $\ell_2$ norm with $||\mathbf{x}||_{2,\epsilon} =  \left(\sum_{i=1}^n x_i^2 + \epsilon\right)^{\frac{1}{2}}$, obtaining a differentiable objective function
\begin{eqnarray} \label{diff_obj}
f_{\epsilon}(\boldsymbol{\beta}) &=& \frac{1}{2}\sum_{i=1}^M\sum_{j=1}^N (y_{ij}-\beta_{ij})^2 + \sum_{i=1}^M \lambda_{1,i}\sum_{j=1}^N ||\beta_{ij}||_{2,\epsilon} \nonumber \\
& & + \sum_{i=1}^M \lambda_{2,i}\sum_{j=2}^N ||\beta_{ij} - \beta_{i,j-1}||_{2,\epsilon} + \sum_{j=2}^N ||\boldsymbol{\lambda}_3 * (\boldsymbol{\beta}_{(j)} - \boldsymbol{\beta}_{(j-1)})||_{2,\epsilon}.
\end{eqnarray}

Adopting an MM framework \cite{LangeOpt}, we want to find a surrogate function $g_{\epsilon}(\boldsymbol{\beta} \mid \boldsymbol{\beta}^{(m)})$ for each iteration $m$ such that $g_{\epsilon}(\boldsymbol{\beta}^{(m)} \mid \boldsymbol{\beta}^{(m)})=f_{\epsilon}(\boldsymbol{\beta}^{(m)})$ and $g_{\epsilon}(\boldsymbol{\beta} \mid \boldsymbol{\beta}^{(m)}) \geq f_{\epsilon}(\boldsymbol{\beta})$ for all $\boldsymbol{\beta}$. At each iteration, then, $\boldsymbol{\beta}^{(m+1)}=\argmin g_{\epsilon}(\boldsymbol{\beta} \mid \boldsymbol{\beta}^{(m)})$.
A majorizing function with the above properties is readily obtained using 
the concavity of square-root function $
||x||_{2,\epsilon} \leq \frac{1}{2||z||_{2,\epsilon}}(x^2 - z^2)
$, and its vector equivalent
$
||\mathbf{x}||_{2,\epsilon} \leq \frac{1}{2||\mathbf{z}||_{2,\epsilon}}(||\mathbf{x}||_{\ell_2}^2 - ||\mathbf{z}||_{\ell_2}^2)
$. The resulting
\begin{eqnarray*}
g_{\epsilon}(\boldsymbol{\beta} \mid \boldsymbol{\beta}^{(m)}) &=&  \frac{1}{2}\sum_{i=1}^M\sum_{j=1}^N (y_{ij}-\beta_{ij})^2 + \sum_{i=1}^M \lambda_{1,i}\sum_{j=1}^N \frac{\beta_{ij}^2}{2||\beta_{ij}^{(m)}||_{2,\epsilon}} \\
& & + \sum_{i=1}^M \lambda_{2,i}\sum_{j=2}^N \frac{(\beta_{ij} - \beta_{i,j-1})^2}{2||\beta_{ij}^{(m)} - \beta_{i,j-1}^{(m)}||_{2,\epsilon}} + \sum_{j=1}^N \frac{||\boldsymbol{\lambda}_3 * (\boldsymbol{\beta}_{(j)} - \boldsymbol{\beta}_{(j-1)})||_{\ell_2}^2}{2||\boldsymbol{\lambda}_3 * (\boldsymbol{\beta}_{(j)}^{(m)} - \boldsymbol{\beta}_{(j-1)}^{(m)})||_{2,\epsilon}} + c^{(m)}
\end{eqnarray*}
can be decomposed in the sum of similar functions of all the row vectors $\boldsymbol{\beta}_i $
\begin{eqnarray}
g_{\epsilon}(\boldsymbol{\beta} \mid \boldsymbol{\beta}^{(m)}) 
& = & \sum_{i=1}^M g_i(\boldsymbol{\beta}_i \mid \boldsymbol{\beta}^{(m)}), \nonumber
\end{eqnarray}
where
\begin{eqnarray} \label{quadratic_form}
g_i(\boldsymbol{\beta}_i \mid \boldsymbol{\beta}^{(m)}) 
& = & \frac{1}{2} \boldsymbol{\beta}_i \mathbf{A}_i^{(m)} \boldsymbol{\beta}_i^T - [\mathbf{b}_i^{(m)}]^T \boldsymbol{\beta}_i^T  + \tilde{c}_i^{(m)}.\label{each}
\end{eqnarray}

Here each $\mathbf{A}_i^{(m)}$ is a tridiagonal symmetric matrix, and $\tilde{c}_i^{(m)}$ is irrelevant constant for optimization purpose.  In view of the strict convexity of the surrogate function, each $\mathbf{A}_i^{(m)}$ is also positive definite. The nonzero entries of $\mathbf{A}_i^{(m)}$ and $\mathbf{b}_i^{(m)}$ ($i=1,\ldots,M$) are listed in the supplementary material. Each of the surrogate functions in (\ref{each}) can be minimized solving the linear system  $\boldsymbol{\beta}_i = [\boldsymbol{\beta}_i^{(m)}]^T[\mathbf{A}_i^{(m)}]^{-1}$ by the Tri-diagonal Matrix (TDM) algorithm \cite{TDMA}.  This results in a per-iteraction computational cost of $O(MN)$. This algorithm is empirically observed to achieve an exponential convergence rate \cite{FL_DPI}, although we do not yet have an analytic proof. In practice, this method scales well with joint analysis of tens to hundreds of samples with measurements at millions of locations, with limitations dictated by memory requirements. For analysis of real data, we suggest one or a group of samples to be analyzed chromosome by chromosome, since a CNV region can never extend beyond one chromosome to another. Actual computation times are shown along with different examples in Section \ref{results}.

\subsection{Stacking observations at different genomic locations}

While copy number is continuously defined across the genome, experimental procedures record data at discrete positions, for which we have used the indexes $j=1,\ldots, N$. In reality, repeated evaluations of the same sample (or related samples) will typically result in measurements at only partially overlapping genomic locations: either because different platforms use different sets of probes, or because missing data my occur at different positions across sequences (consider for example, mBAF and LRR from the same experiment on one subject: the mBAF signal will be defined on a subset of the locations where LRR is).

Let $S$ indicate the union of all genomic positions where some measurement is available among the $M$ signals under study. And let $S_i$ be the subset of locations with measurements in sequence $i$. We reconstruct $\beta_{ij}$ for all $j \in S$. When $j \notin S_i$, $\beta_{ij}$ will be determined simply on the basis of the neighboring datapoints, relying on the regularizations introduced in (\ref{mmodel}). The goodness-of-fit portion of the objective function is therefore redefined as 
\begin{equation} \label{model_miss}
\frac{1}{2}\sum_{i=1}^M\sum_{j=1}^N (\delta_{ij}y_{ij}-\delta_{ij}\beta_{ij})^2 \;\;\;\; \mbox{with}\;\;\;\;
\delta_{ij} = 
\left \{ \begin{array}{ll} 
1, & \mbox{if } j \in S_i,\\
0, & \mbox{otherwise}.
\end{array}  \right.
\end{equation}
The  MM strategy  previously described applies with slight modifications of the matrix $\mathbf{A}_i^{(m)}$ (see the supplementary material).

The attentive reader would have noted that $y_{ij}$ with $j\notin S_i$ can be considered as missing data, and an evaluation of the characteristics of this missingness is appropriate. In general, $y_{ij}$ cannot be considered missing at random. The most important   example is  the case of mBAF, where homozygous markers result in missing values. Now, homozygosity is more common when copy number is equal to 1 than when copy number is equal to 2 and, therefore, there is potentially more information on $\beta_{ij}$ to be extracted from the signals than the one we will capture with the proposed methodology. On the other hand, it does appear that the approach outlined does not increase false positive: operationally, then, it can be considered as an improvement over segmentation based on LRR only, even if in theory, it does not completely use the information on BAF. It is also relevant to note that, in reality, most of the information on deletion is obtained through LRR, and BAF is really carrying additional information in case of duplications (where the changes in LRR are limited due to saturation effects).

\subsection{Choice of tuning constants and segmentation}
One of the limitations of penalization procedures is that a value for the tuning parameters needs to be set and clear guidelines are not always available. Path methods that obtain a solution of the optimization problem (\ref{mmodel}) for every value of tuning parameters can be attractive, but recent algorithmic advances \cite{GroupFL_new, path_ryan, path_hua_ken} remain impractical for problems of the size of ours. 
A number of recent publications obtain optimal values of penalty parameters under a series of conditions \cite{bickel, bunea, Dantzig_selector, Donoho}: we rely upon them to propose the following strategy consisting of obtaining a solution of  (\ref{mmodel}) for reasonably liberal values of the tuning parameters, followed by a sequence-by-sequence hard thresholding of the detected jumps with a data-adaptive threshold.  

We have found the following guidelines to be useful in choosing penalty parameter values:
\begin{eqnarray} \label{tuning_par}
\lambda_{1,i} &=& c_1 \hat{\sigma}_{i}, \nonumber \\
\lambda_{2,i} &=& \rho(p) c_2 \hat{\sigma}_{i}\sqrt{\log N}, \label{penalty} \\
\lambda_{3,i} &=& [1-\rho(p)]c_3 \hat{\sigma}_{i}\sqrt{pM}\sqrt{\log N},\nonumber
\end{eqnarray}
for $i=1,\ldots,M$,  where $\hat{\sigma}_{i}$ is a robust estimate of standard deviation of $\mathbf{y}_i$, $p$ is roughly the proportion of the $M$ sequences we anticipate to carry CNVs, and $c_1$, $c_2$ and $c_3$ are positive multipliers adjusted in consideration of different signal-to-noise ratios and CNV sizes.

While a more rigorous justification is provided in the supplementary material, we start by underscoring some of the characteristics of this proposal.
\begin{itemize}
\item The sequence-specific penalizing parameters are proportional to an estimate of the standard deviation of the sequence signal: that is, proviso an initial normalization, the same penalties would be used across all signals.
\item The tuning parameter for the total variation (fused lasso) and the Euclidean  (group fused lasso) penalties on the jumps depend on $\sqrt{\log{N}}$, where $N$ is the possible number of jumps. This has a ``multiple comparison controlling'' effect and resembles rates that have been proven optimal under various sparse scenarios \cite{bickel, bunea, Dantzig_selector, Donoho}.
This term does not appear in the expression of $\lambda_1$, as the lasso penalty can be understood as providing a soft thresholding of the solution of (\ref{mmodel}) when $\lambda_1=0$: given the penalization due to $\lambda_2$ and $\lambda_3$, this object will have much smaller dimensionality than $N$.
\item The group penalty depends on $\sqrt{M}$, where $M$ is the number of grouped sequences, as in the original proposal \cite{GroupLasso}.
\item The relative weight of the fused-lasso and group-fused-lasso penalties is regulated by $\rho$, which depends on $p$, the proportion of the $M$ sequences expected to carry the same CNV. For example, if $M=2$ and the two sequences are LRR and BAF from the same individual, we anticipate $p=1$ with $\rho=0$, enforcing jumps at identical places in the two signals. At the other extreme, for completely unrelated sequences, $p=0$ and $\rho=1$. 
\end{itemize}

The standard deviation $\hat{\sigma}_i$ can be estimated robustly as follows. Let $\Delta_{ij}=y_{i,j+1}-y_{i,j}$, for $j=1,\ldots,N-1$, be the one-order difference of adjacent $y_{ij}$ for sequence $i$. Then most $\mbox{Var}(\Delta_{ij}) = 2\sigma_i^2$ except those bridging real change points, so we can take
$$
\hat{\sigma}_i = \widehat{SD}(\boldsymbol{\Delta}_i)/\sqrt{2},
$$
where $\widehat{SD}(\boldsymbol{\Delta}_i) = \mbox{Standard Deviation}(\boldsymbol{\Delta}_i)$ or $\widehat{SD}(\boldsymbol{\Delta}_i) = \mbox{Median Absolute Deiviation}(\boldsymbol{\Delta}_i)$ for $\boldsymbol{\Delta}_i = \{\Delta_{i,1},\ldots, \Delta_{i,N-1}\}$. 

As mentioned before, the exact values of the penalty parameters should be adjusted depending on the expectations of signal strengths. Following the approach in \cite{FL_AOS}, one can approximate the bias induced by each of the penalties and hence work backwards in terms of acceptable levels. As detailed in the supplementary material, 
\begin{eqnarray*}
\mbox{Bias}(\lambda_1)& \approx & \lambda_1 \\
\mbox{Bias}(\lambda_2)& \approx & \lambda_2/\mbox{Length of segment}\\
\mbox{Bias}(\lambda_3)& \approx & \lambda_3/(\mbox{Length of segment} \times \sqrt{\mbox{\# sequences sharing segment}})
\end{eqnarray*}

Following again the approach in \cite{FL_AOS}, one can show that under some relatively strong assumptions, the choices in (\ref{penalty}) lead to a consistent behavior as $N \to \infty$ and $M$ stays bounded (see the supplementary material). Despite the fact that $N$ is indeed large in our studies, it is not clear that we can assume it to be in the asymptotic regime. As finer scale measurements become available, scientists desire to investigate CNV of decreasing length: the CNVs we are interested in discovering are often covered by a small number of probes.  Furthermore we have often little information on the sizes and frequencies of CNV. In this context, we find it advisable to rely on a two-stage strategy:
\begin{enumerate}
\item Sequences are jointly segmented minimizing (\ref{mmodel}) for a relatively lax choice of the penalty parameters.
\item Jumps are further thresholded on the basis of a data-driven cut-off.
\end{enumerate}
Step 2 allows us to be adaptive to the signal strength and can be carried on with multiple methods. For example, one can adopt the modified Bayesian Information Criteria (mBIC) \cite{mBIC}. For sequence $i$, the jumps are sorted as $\{\hat{d}_{i(1)},\ldots,\hat{d}_{i(N-1)}\}$ in the descending order of their absolute values. And then we choose the first $\hat{k}$ change points where $\hat{k}$ is given by
$$
\hat{k} = \argmax_{k} \mbox{mBIC}(k).
$$
In data analysis, we often apply an even simpler procedure where the threshold for jumps is defined as a fraction of the maximal jump size observed for every sequence. Specifically, for sequence $i$, let  $\hat{D}_i=\max_{2 \leq j \leq N}\{|\hat{d}_{ij}|\}$, where $\hat{d}_{ij} = \hat{\beta}_{ij}-\hat{\beta}_{i,j-1}$, be the largest observed jump for sequence $i$.  Then we define
$$
\gamma_i = \max\{a\hat{\sigma}_i,\min\{\hat{D}_i,b\hat{\sigma}_i\}\}, \quad \mbox{for } a<b,
$$
as a ``ruler'' reflecting the scale of a possible real jump size, taking $c\gamma_i$ as the cut-off in removal of most small jumps. In all analyses for this paper, we fix $a=1$, $b=5$ and $c=0.2$. In our experience, this heuristic procedure works well for both tumor and normal tissue CNV data. 

\subsection{Calling Procedure}

Even if this is not the focus of our proposal,  in order to compare the performance of our segmentation algorithm with HMM approaches, it becomes necessary to distinguish acquisitions from losses of copy number. 
While the same segmentation algorithm can be applied to a wide range of data sets, calling procedures depend more closely on the specific technology used to carry out the experiments. Since our data analysis relies on Illumina genotyping arrays, we limit ourselves to this platform, and briefly describe the calling procedure we adopt in Section \ref{results}. 

Analyzing one subject at the time, each segment with constant mean is assigned to one of five possible copy number states ($c=0,1,2,3,4$). Let 
$R$ collect the indexes of all SNPs comprising one segment and let $(\mathbf{x}_R,\mathbf{y}_R)=\{(x_j,y_j),j \in R\}$ be the vectors of values for BAF and LRR in the segment. On the basis of typical pattern for BAF and LRR in the different copy number states (see \cite{QuantiSNP, us, PennCNV}), we can write log-likelihood ratio
\begin{equation} \label{lr}
\mbox{LR}(c) = \log\frac{L_{\mbox{\scriptsize BAF}}(\mathbf{x}_R;c)}{L_{\mbox{\scriptsize BAF}}(\mathbf{x}_R;2)} + \log\frac{L_{\mbox{\scriptsize LRR}}(\mathbf{y}_R;c)}{L_{\mbox{\scriptsize LRR}}(\mathbf{y}_R;2)},\quad c=0,1,3,4,
\end{equation}
explicitly defined in the supplementary material. 
 Segment $R$ is assigned a CNV state $\hat{c}$ that maximize $\mbox{LR}(c)$, only if $\mbox{LR}(\hat{c}) > r_1$, where $r_1$ is a pre-specified cut-off.

As noted in \cite{Nancy_Biometrika10},  the LRR data for a segment with $c=2$, ideally normalized to have mean $0$, often has a small non-zero mean, due to experimental artifacts. If the number of SNPs in $R$ is sufficiently large, a log-likelihood-ratio criterion as the above would result in the erroneous identification of a copy number different from 2. To avoid this, we also require that the size of the absolute difference of the mean of LRR from zero be larger than a threshold $|\bar{y}_{R}|> r_2\sigma$.

\section{Results}  \label{results}
We report the results of the analysis of two simulated and two real data sets, which overall exemplify the variety of situations where joint segmentation of multiple sequences is attractive, as described in the introduction. In all cases, we compare the performance of the proposed procedure with a set of relevant, often specialized, algorithms. The penalized estimation method we put forward in this manuscript shows competitive performance in all cases and often a substantial computational advantage. Its versatility and speed make it a very convenient tool for initial exploration.
To calibrate the run times reported in what follows, it is relevant to know that all our analyses were run on a Mac OS X (10.6.7) machine with 2.93 GHz Intel Core 2 Duo and 4 GB 1067 MHz DDR3 memory.

\subsection{Simulated CNV in normal samples}
We consider one of the simulated data sets described in \cite{FL_DPI}: relatively short deletion and duplication (300 comprising 5,10, 20, 30, 40, 50 SNPs each) are inserted in the middle of 13000 SNPs long sequences, using a combination of male and female X chromosome data from Illumina HumanHap550 array, appropriately pre-processed to avoid biases (these steps included a scrambling of SNP positions, so to avoid long-range signal fluctuation). This setting  mimics the small rare CNVs possibly occurring in the genome of normal individuals: in our main analysis, therefore, we process one individual at the time, reflecting the typical level of information available to scientists in these contexts. HMM methods, like PennCNV, are expected to be the most effective in this problem; segmentation methods like CBS are closer to our own and therefore also make an interesting comparison. As repeatedly discussed, Illumina platform produces two signals for one subject: LRR and BAF. A segmentation method that can process one signal at the time would give its best results using LRR, which carries most of the information. Given this background, we compare four methods: PennCNV, CBS on LRR, fused lasso on LRR only, and group fused lasso on LRR and mBAF. The implementations we use are those reflected in the software packages: PennCNV (version 2010May01), R package DNAcopy for CBS (version 1.24.0) \cite{fastCBS} and our own R package Piet (version 0.1.0). Tuning parameters for PennCNV and CBS are set at the default values; the fused lasso implementation corresponds to $\lambda_1=0.1$, $\lambda_2=2\times\sqrt{13000}$, and  $\lambda_3=0$ and the group fused lasso to $\lambda_1=0.1$, $\lambda_2=0$, and $\lambda_3=2\times\sqrt{13000}$. To call deletion and duplication with CBS and the two fused-lasso approaches, we use both LRR and BAF data (before transformed to mBAF) with the following cut-off values: $r_1=10$ and $r_2=1 (1.5)$  for duplication (deletion). Performance is evaluated  by the same indexes we used in \cite{FL_DPI}: true positive rate (TPR or sensitivity) and false discovery rate (FDR), all  defined on a per SNP basis. Results are summarized in Table \ref{NormalSimulation}.

Not surprisingly, all algorithms perform similarly well for larger deletions/duplications and it is mainly for variants that involve $\leq 10$ SNPs that differences are visible. Algorithms that rely only on LRR (as CBS and fused lasso) underperform in the detection of small duplications (comparison is particularly easy for duplications of size 10 SNP, where the selected parameter values lead to similar FDRs in the three segmentation methods). The group fused lasso can almost entirely recover  the performance of PennCNV and outperforms CBS in this context.

\begin{table}[htb]
\begin{center}
\caption{Detection accuracy (as percentage of SNPs) and computation times for PennCNV, CBS, Fused Lasso and Group Fused Lasso on a simulated set of CNVs in normal samples. Overall accuracy are calculated  pooling all sequences with a given type of CNVs. The average (and standard deviation) of the number of seconds required for the analysis of one sequence is reported.}
\label{NormalSimulation}
\hspace*{-.8cm}
{\footnotesize
\begin{tabular}{cl|cc|cc|cc|cc}
\hline
CNV & CNV & \multicolumn{2}{c|}{PennCNV} & \multicolumn{2}{c|}{CBS}  & \multicolumn{2}{c|}{Fused Lasso} & \multicolumn{2}{c}{Group Fused Lasso} \\
\cline{3-4}\cline{5-6}\cline{7-8}\cline{9-10}
Size & Type & TPR & FDR &  TPR & FDR & TPR & FDR & TPR & FDR \\
\hline
 5  & Deletion  & 83.80 & 4.92 & 78.20 & 0.68 & 63.93 & 1.74 & 64.27 & 1.83 \\
     & Duplication & 58.53 & 4.67 & 11.67 & 10.26 & 20.00 & 37.76 & 39.87 & 14.33 \\
10 & Deletion  & 95.03 & 1.45 & 88.37 & 0.56 & 88.50 & 0.60 & 88.87 &  0.56 \\
     & Duplication & 93.43 & 0.78 & 56.50 & 4.40 & 83.90 & 12.60 & 91.60 & 3.85 \\
20 & Deletion  & 94.63 & 0.58 & 90.50 & 0.39 & 90.80 & 0.47 & 90.83 & 0.47 \\
     & Duplication & 96.13 & 0.92 & 86.22 & 3.58 & 92.77 & 4.95 & 94.98 & 2.13 \\
30 & Deletion  & 94.57 & 0.28 & 93.30 & 0.29 & 89.38 & 0.52 & 89.77 & 0.53 \\
     & Duplication & 96.09 & 0.05 & 90.77 & 1.61 & 94.32 & 1.78 & 94.98 & 1.29 \\
40 & Deletion  & 97.83 & 0.59 & 97.58 & 0.09 & 97.28 & 0.19 & 97.28 & 0.19 \\
     & Duplication & 94.61 & 0.46 & 92.77 & 0.98 & 93.94 & 1.15 & 94.63 & 0.75 \\
50 & Deletion  & 94.33 & 0.07 & 92.76 & 0.04 & 90.47 & 0.11 & 90.48 & 0.11 \\
     & Duplication & 94.50 & 0.09 & 93.81 & 0.74 & 93.11 & 0.79 & 93.64 & 0.49 \\
\hline
\multicolumn{2}{c|}{Overall Deletion} & 95.02 & 0.55 & 93.06 & 0.19 & 91.08 & 0.33 &  91.19 & 0.34 \\
\multicolumn{2}{c|}{Overall Duplication} & 93.82 & 0.44 & 86.92 & 1.55 & 90.56 & 2.85 &  92.46 & 1.38 \\
\multicolumn{2}{c|}{Overall} & 94.42 & 0.49 & 89.99 & 0.85 & 90.82 & 1.60 &  91.83 & 0.87 \\
\hline
\multicolumn{2}{c|}{Time (sec.)} & \multicolumn{2}{c|}{0.48 (0.01)} & \multicolumn{2}{c|}{0.78 (0.69)} & \multicolumn{2}{c|}{0.22 (0.13)} & \multicolumn{2}{c}{0.28 (0.05)} \\
\hline
\end{tabular}}
\end{center}
\end{table}

For curiosity, we analyzed all sequences simultaneously. While this represents an unrealistic amount of prior information, it allows us to evaluate the possible gain of joint analysis: FDR practically become 0 ($<$0.02\%) for all CNV sizes, but power increases only for CNV including less than 10 SNPs.

Finally, it is useful to compare running times. Summary statistics of the per sample time are reported in Table \ref{NormalSimulation}: while all algorithms are rather fast, the two implementations of the fused lasso are dominating.

\subsection{A simulated tumor data set}
To explore the challenges presented by  tumor data, we rely on a data set created by \cite{Staaf08}, with the specific goal of studying the effect of contamination between normal and cancer cells. The HapMap sample NA06991, genotyped on Illumina HumanHap550 array, was used to simulate a cancer cell line, by inserting  a total of $10$ structure variation regions, including one-copy losses, one-copy gains, and copy neutral loss-of-hetrozygosity (CN-LOH) (see Supplementary Table \ref{CNVdescription}). The signal from this artificial ``tumor" sample was then contaminated in silico with that of the original ``normal" sample, resulting in $21$ data sets, with a percentage of normal cells ranging from $0\%$ to $100\%$. Note that most simulated CNV or CN-LOH regions are very large---some spanning an entire chromosome---and the challenge in detection is really due to the contamination levels. 
 
For ease of comparison, we evaluate the accuracy of calling procedures as in the original reference \cite{Staaf08}: sensitivity is measured for each variant region as the percentage of heterozygous SNPs that are assigned the correct copy number; and specificity is the percentage of originally heterozygous SNPs in unperturbed regions that are assigned CN=2. We compare the performance of GFL to BAFsegmentation \cite{Staaf08} and PSCN \cite{PSCN} representing, respectively, a version of segmentation and HMM approaches specifically developed to deal with contaminated tumor samples (both these algorithms have been tested with success on this simulated data set).
 
Following other analyses, we do not pre-process the data prior to CNV detection. BAFsegmentation and PSCN were run using recommended parameter values. For each of the diluted data sets, we applied the GFL model on each chromosome at one time using both LRR and mBAF, whose standard deviations are normalized to $1$. Tuning constants are set to $\lambda_1=0$, $\lambda_2=0.5\times3\times\sqrt{\log N}$, and $\lambda_3=0.5\times3\times\sqrt{\log N}$, varying specifically for chromosome interrogated by $N$ SNPs. The change points resulting from hard segmentation on LRR and mBAF are combined to make a finer segmentation of the genome. Finally, we adopt the same calling procedure described by \cite{Staaf08}. For ease of comparison with PSCN, only analysis of simulated tumor data are reported, even if  BAFsegmentation and GFL would gain from using the genotype of normal cell in defining mBAF.

\begin{figure}[htb]
\centering
\includegraphics[width=1\textwidth]{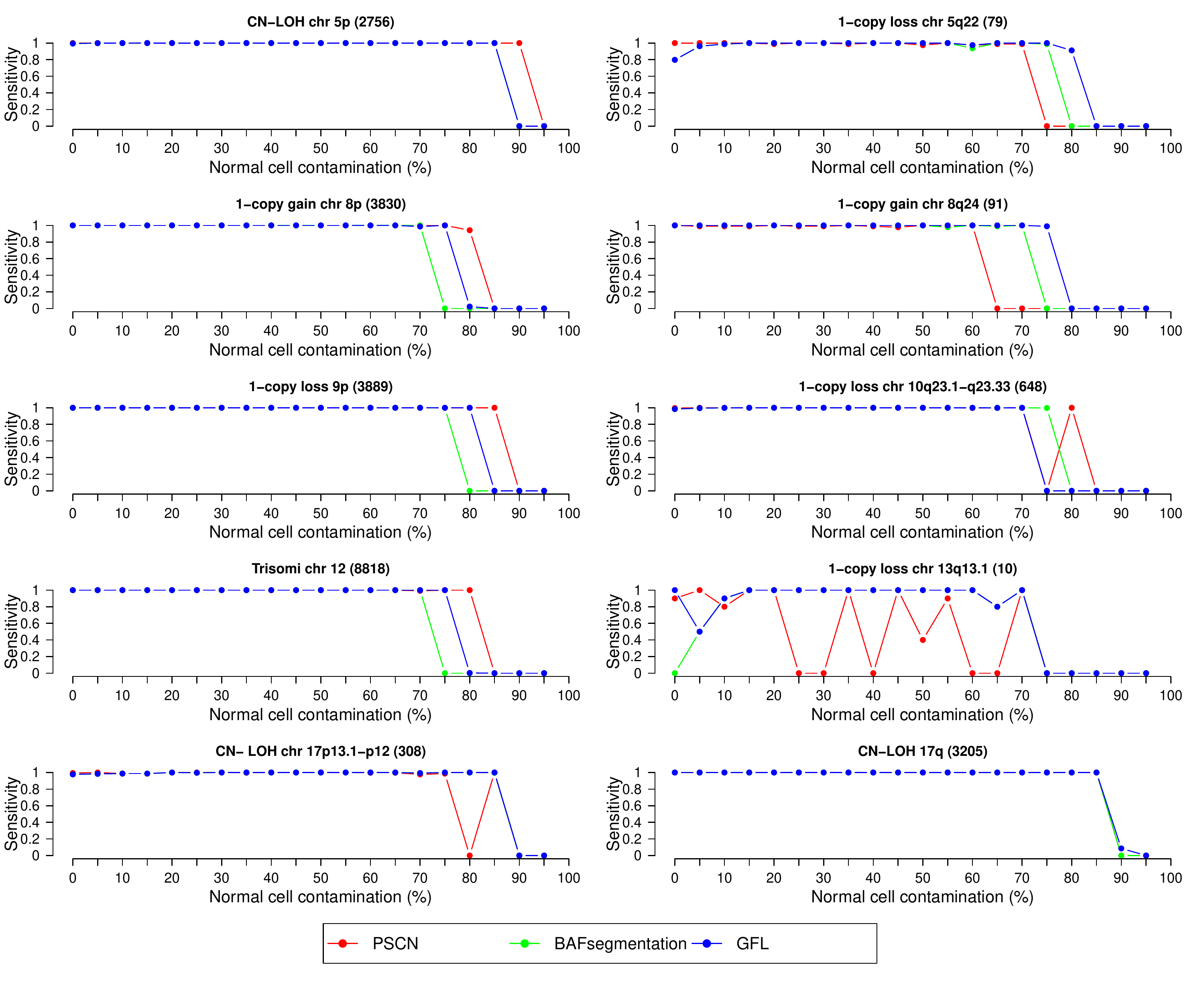}
\caption{Sensitivity as function of percentage contamination by normal cells in the 10 different simulated CNV regions. Sensitivity is not defined at 100\% contamination.}
\label{sensitivity_tumor}
\end{figure}

\begin{figure}[htb]
\centering
\includegraphics[width=0.8\textwidth]{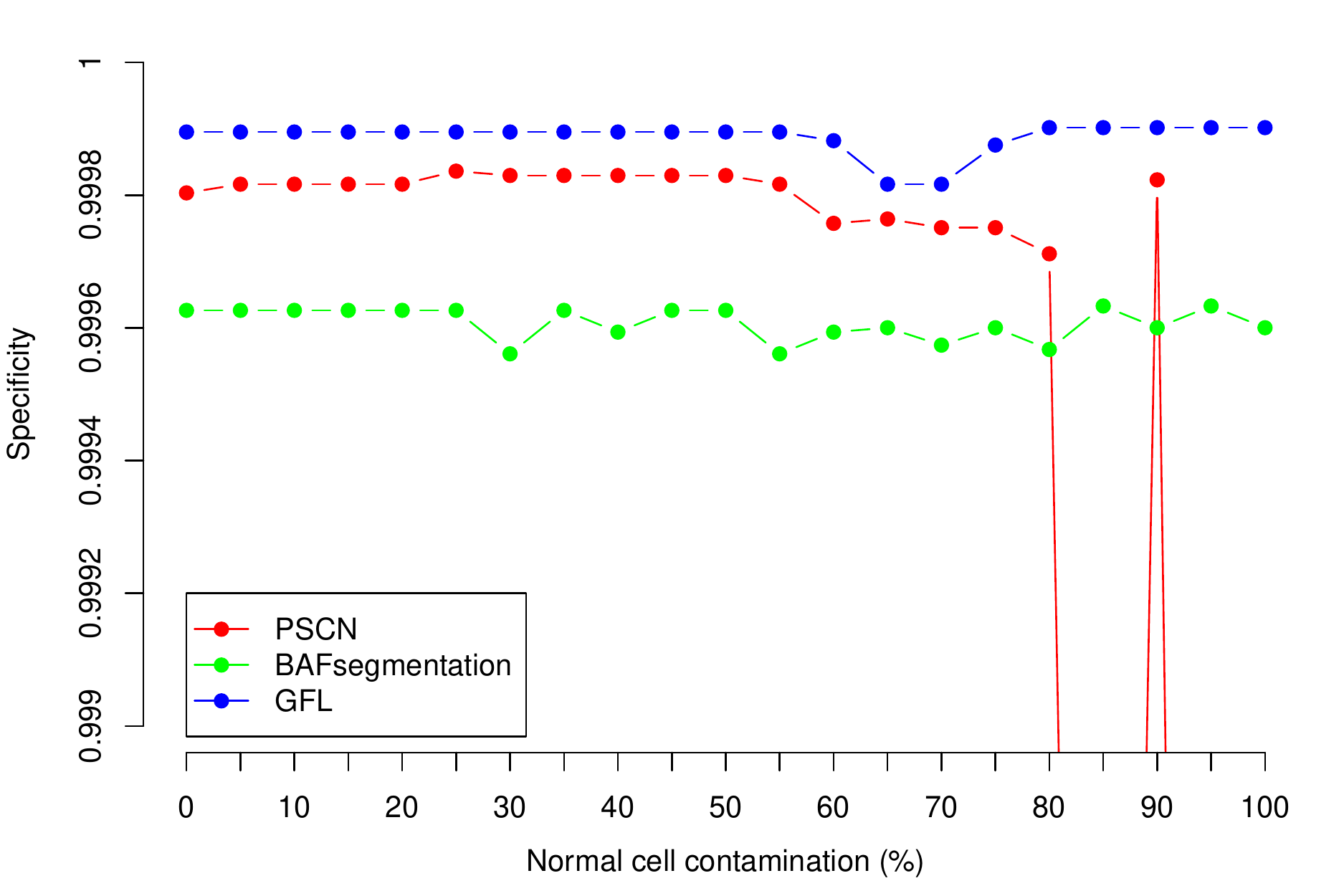}
\caption{Specificity as function of percentage contamination by normal cells. Note that \cite{PSCN} reports better perfomance of PSCN in correspondence of contamination levels 85\% , 95\% and 100\%.}
\label{specificity_tumor}
\end{figure}

Figure \ref{sensitivity_tumor} summarizes the sensitivity of each method, as a function of percentage of normal cell in the sample. Sensitivity is calculated for each of the $10$ regions separately. All three methods work reasonably well under a wide range of percentages of normal cell contamination (in $5$ out of the $10$ regions, GFL appears to lead to best results, while in the other $5$ PSCN does). The CNV region that comprises the smallest amount of SNP is the hemizygous loss on Chromosome 13: in this case GFL in our hands behaved in the most stable manner. GLF outperforms the two comparison methods in terms of specificity (Figure \ref{specificity_tumor}): while the specificity values might appear very high in any case, this is somewhat of an artifact due to the adopted definition of this index. It is relevant to note that the performance of PSCN in our hands does not correspond to the published one \cite{PSCN}. While we tried our best to set the parameter values, we have not succeeded in replicating the authors' original results, which should be considered in the interest of fairness. 

PSCN, like GFL, is implemented in R with some computationally intensive subroutines coded in C. BAFsegmentation relies its segmentation part on the R package DNAcopy, whose core algorithms are implemented in C and Fortran, and it is wrapped in Perl.  A comparison of run times indicate that GLF and BAFsegmentation are comparable, while PSCN is fifty times slower than GFL (see Supplementary Table \ref{tumor_time}). 

\subsection{One sample assayed with multiple replicates and multiple platforms}

We use the data from a study \cite{Nature_Biotech_data} assessing the performance of different array platforms and CNV calling methods to illustrate the advantages of joint analysis of multiple measurements on the same subject. DNA from four individuals was analyzed in triplicate on each of 5 platforms: Affymetrix 6.0, Illumina 1M,  660W, Omni1-Quad (O1Q) and Omni2.5-Quad (O2Q) (among others \cite{Nature_Biotech_data}). We use the results on the first three to define ``true'' copy numbers and try to reconstruct them using data from O1Q and O2Q. The nine ``reference" experiments were analyzed with 4 or 5 CNV calling algorithms (see \cite{Nature_Biotech_data}) and a CNV was identified using majority votes: 
consistent evidence was required from at least 2 analysis tools, on at least 2 platforms, and in at least 2 replicates (see Supplementary Table \ref{sample_info}). Here CNVs detected in two replicates/algorithms/platforms are regarded as the same CNV and collapse down to one CNV with the outmost boundaries when they overlap with each other.

The test experiments are based on 1,020,596 and 2,390,395 SNPs on autosomes after some quality control, at a total of 2,657,077 unique loci. Since our focus here is to investigate how to best analyze multiple signals on the same subject, rather than on the specific properties of any CNV calling method, we carry out all the analyses using different settings of GFL in segmentation while keeping the same CNV calling and summarizing procedure. All segmentation is done on LRR only while calling procedure uses both LRR and BAF (with cut-off $r_1=10$ and $r_2=1$). Here we compare three segmentation settings to analyze these 6 experiments per subject (see Supplementary Table \ref{multipltfm_data} for more details about tuning parameters):
\begin{enumerate}
\item The signals from the three technical replicates with one platform are averaged and then segmented and subject to calling procedure separately. The final CNV list is the union of CNV calls from the two platforms.
\item The signals from the three technical replicates with one platform are each segmented and subject to calling procedure separately. A majority vote is used to summarize CNV result for each platform: a CNV needs to be called in at least two replicates out of three. The final CNV list is the union of the two platforms' results.
\item The signals from the three technical replicates of both platforms (6 LRR sequences) are segmented jointly. Calling procedure is still done on each replicate separately, and the same majority vote is used to summarize CNV result for each platform. Again, the final CNV list is the union of the two platforms' results.
\end{enumerate}
To benchmark the result of joint analysis we use MPCBS \cite{Nancy_Bioinformatics10}, a segmentation method,  specifically designed for multi-platform CNV analysis. The segments output from MPCBS are proceeded to the same calling, majority voting, and summarizing procedure.

\begin{table}[htb]
\begin{center}
\caption{Number of CNVs detected (Det.) and overlapping (Ovlp.) with reference results as well as average computation time for four samples under different analyses.}
\label{multipltfm_res}
\hspace*{-0.8cm}
{\footnotesize
\begin{tabular}{c|cc|cc|cc|cc|c}
\hline
                    & \multicolumn{2}{c|}{NA15510} &  \multicolumn{2}{c|}{NA18517} & \multicolumn{2}{c|}{NA18576} &  \multicolumn{2}{c|}{NA18980} &  \\
\cline{2-9}
\hline
Analysis      &   \# Det. & \# Ovlp.        &  \# Det. & \# Ovlp.          &  \# Det. & \# Ovlp         &  \# Det. & \# Ovlp  & Time (min.) \\
\hline
Analysis 1    & 170    & 38                     & 144  & 34                      &  160  & 25                    &  145  & 22              & 1.2 \\
Analysis 2    & 102    & 36                     & 109  & 33                      &  93  & 25                      &  91  & 20                 & 3.7 \\
Analysis 3    &  80     & 38                     &  82   & 32                      &  69  & 25                      &  56 & 15                  & 8.5 \\
MPCBS        &  98      & 34                    &  88   & 28                      &  59  & 18                      & 68   & 21                  & 313.9 \\
\hline
\end{tabular}}
\end{center}
\end{table}

Table \ref{multipltfm_res} presents the results: averaging results from different technical replicates leads to loss of power, while joint analysis of all the signals leads to the most effective performance. GFL joint analysis leads to results comparable to those of MPCBS, but it is at least 30 times faster than the competing method.

\subsection{Multiple related samples assayed with the same platform} \label{BP_study}
In the context of a study of the genetic basis of bipolar disorder, the Illumina Omni2.5-Quad chip was used to genotype 455 individuals from 11 Columbian and 13 Costa Rican pedigrees. We use this data set to explore the advantages of a joint segmentation of related individuals. In absence of a reference evaluation of CNV status in these samples, we rely on two indirect methods to assess the quality of the predicted CNVs.
We used the collection of CNVs observed in HapMap Phase III \cite{HapMapCNV} to compile a list of 426 copy number polymorphisms (selecting all those CNVs with frequency $\geq 0.05$ in pooled samples from $11$ populations) and assumed that if we identify in our sample a CNV corresponding to one of these regions, we should consider it  a true positive.
For the purposes of this analysis we considered a detected CNV to correspond to one identified in HapMap if there was any overlap between the two regions.

Another indirect measure of the quality of CNV calls derives from the amount of Mendelian errors encountered in the pedigrees when we consider the CNV as a segregating site. De novo CNVs are certainly a possibility, and in their case Mendelian errors are to be expected. However, when the CNV in question is a common one (already identified in HapMap), it is reasonable to expect that it segregate in the pedigrees as any regular polymorphism.  We selected a very common deletion on Chromosome 8 (HapMap reports overall frequency $>0.4$ in $11$ populations) and compared different CNV calling procedures on the basis of how many Mendelian errors they generate.

As mentioned before, PennCNV represents a state-of-the-art HMM method for the analysis of normal samples and, therefore, we included it in our comparisons. However, the parameters of the underlying HMM algorithm had not been tuned on the Omni2.5-Quad at the time of writing, resulting in sub-standard performance. Segmentation methods are less dependent on parameter optimization; hence, GFL analysis of LRR and BAF one subject at a time can provide a better indication of the potential of single-sample methods. We considered two multiple-sample algorithms: GFL  and MSSCAN \cite{Nancy_Biometrika10}, both applied on LRR with group defined by pedigree memberships. (While a trio-mode is available in PennCNV \cite{Wang_PCNV_joint}, this  does not adapt to the structure of our families.) A final qualification is in order. While the authors of MSSCAN kindly shared with us a beta-version of their software, we find it not to be robust. Indeed, we were unable to use it to segment the entire genome. However, we successfully used it to segment Chromosome 8, so that we could include MSSCAN in the comparison based on Mendelian error rates.

Prior to analysis, the data was normalized using the GC-content correction implemented in PennCNV \cite{gcadj}. For individual analysis, the GFL parameters were 
 $\lambda_1=0.1$, $\lambda_2=0$, and $\lambda_3=2\times\sqrt{\log N}$, where $N$ is the number of SNPs deployed on each chromosome; for pedigree analysis, the GFL parameters were $\lambda_1=0.1$, $\lambda_2=0.5\times2\times\sqrt{\log N}$, and $\lambda_3=0.5\times2\times\sqrt{0.3M}\times\sqrt{\log N}$, where $M$ is the number of individuals in each pedigree. 
For MSSCAN,   CNV size is constraint to be less than 200 SNPs and the maximum number of change points is set as 50.
 The calling procedure with $r_1=10$ and $r_2=1$ was applied to both the GFL and MSSCAN results.
 
Table \ref{CNPR_overlap} summarized the total number of copy number polymorphisms (CNPs) identified in our sample by different approaches and their overlap with known CNPs from HapMap. For the purpose of this comparison we considered as a CNP a variant with frequency at least 10\% in our sample.
All analysis modes of GFL agree more with HapMap list than PennCNV in the sense of percentage of overlap. It is also clear that GFL-pedigree analysis achieves larger overlap with HapMap data than GFL-individual analysis. The time cost per sample for pedigree is reasonable and scales well with the increment of sample size. 

\begin{landscape}
\begin{figure}[htb]
\centering
\includegraphics[width=1.4\textwidth]{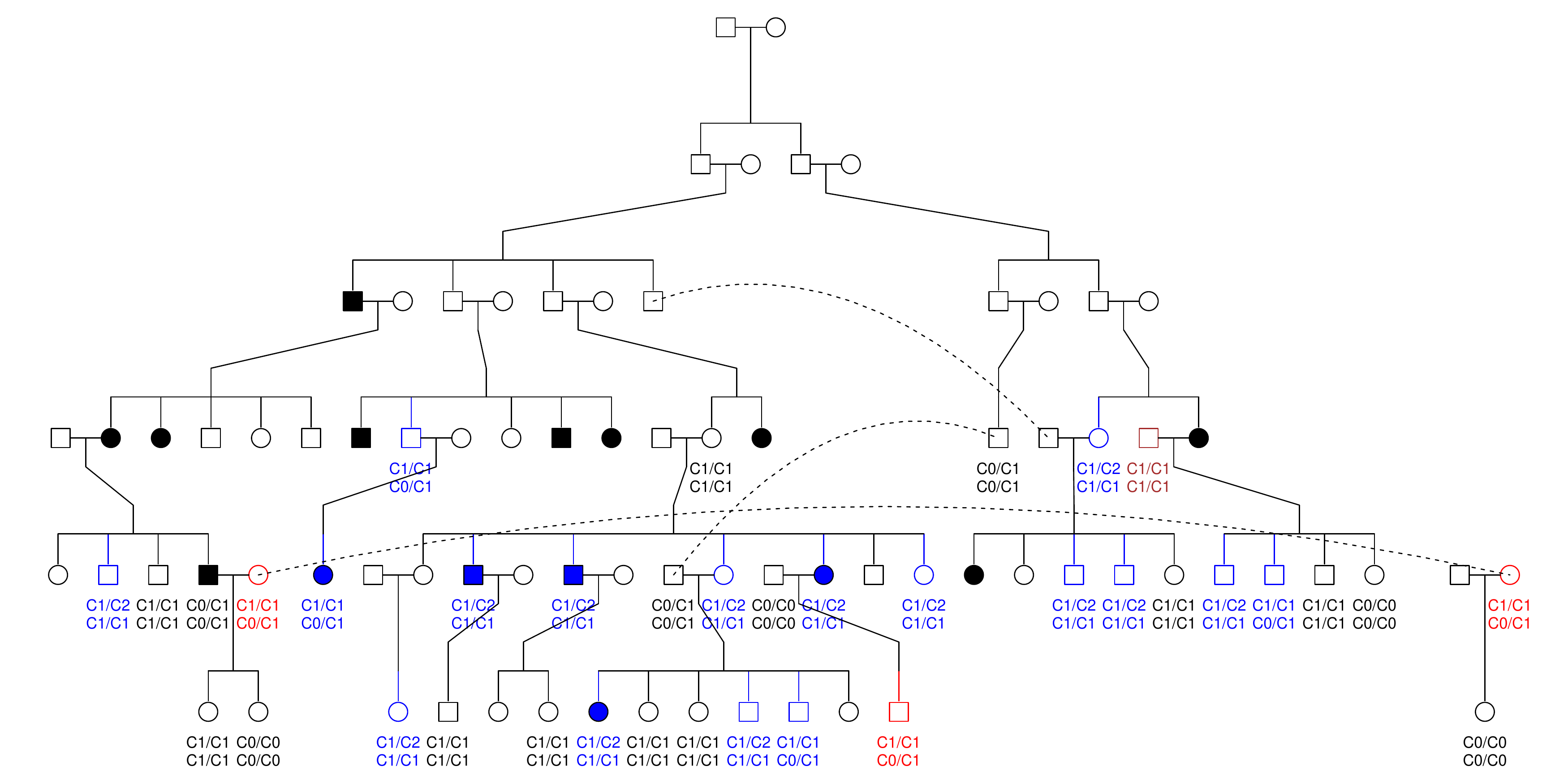}
\caption{Costa Rican pedigree CR004: Circles and squares, filled and empty, have the usual interpretation in pedigree drawing. Beneath each individual, from top to bottom, are CNV genotypes by PennCNV and by GFL. The subjects for whom PennCNV and GLF infer different CNV genotypes are highlighted in red and blue. Red indicates cases where the PennCNV genotype results in Mendelian error, while blue is for subjects where both genotypes are compatible with the rest of the family. Brown indicates a member for whom both PennCNV and GFL genotypes result in Mendelian error.}
\label{CR004_PCNV_GFL}
\end{figure}
\end{landscape}

Table \ref{CNPR_chr8} summarizes the results of our investigation of a 154kb CNP region on Chromosome 8p (from 39,351,896 to 39,506,122 on NCBI Build 36 coordinate).  All methods but PennCNV show detected deletions only; this coincides with the observation from HapMap data. We used option {\em Mistyping} of Mendel (version 11.0) \cite{mendel, mendel_mistyping} to detect Mendelian errors. Joint segmentation methods discover more hemizygous deletions than individual analysis, resulting in fewer Mendelian errors.  MSSCAN discovers the largest number of  hemizygous deletions. Figure \ref{CR004_PCNV_GFL} shows an example of large pedigree, where 3 out of 4 Mendelian errors are removed by joint analysis.
  
\begin{table}[htb]
\begin{center}
\caption{The number of detected CNP regions with  frequency $\geq 0.1$ in our sample by different methods and their overlap with a list of CNP regions compiled from HapMap data. Computation time (in minute) is per sample.}
\label{CNPR_overlap}
\begin{tabular}{l|ccc|cc|c}
\hline
Method & \# detected CNVR & \# Overlap & \% Overlap & Time (min.)\\
\hline
PennCNV                                      & 189 & 63   & 33.33\%     & 3.44 \\
GFL-Individual (LRR+BAF)       & 95   & 50   & 52.63\%     & 3.90 \\
GFL-Pedigree  (LRR)                 & 106 & 62   & 58.49\%     & 1.57 \\
\hline
\end{tabular}
\end{center}
\end{table}

\begin{table}[htb]
\begin{center}
\caption{Detected copy numbers in a common deletion  on Chromosome 8. Across the various algorithms, subjects are assigned to one of 4 types of copy number: for each algorithm, we report the total numbers of CN$\neq 2$ identified; the total number of ``core'' families with Mendelian errors; and the average computation time (in minute) per sample for the analysis of Chromosome 8.}
\label{CNPR_chr8}
\begin{tabular}{l|ccc|c|c}
\hline
Method & \# CN=0 & \# CN=1 & \# CN=3 & \# families with Mendelian errors & Time (min.)\\
\hline
PennCNV                  & 125 & 39   & 102 &  35 & 0.19 \\
GFL-Individual          & 123 & 97   & 0     & 20 & 0.21 \\
GFL-Pedigree           & 123 & 137 & 0      & 15 & 0.09 \\
MSSCAN-Pedigree & 123 & 154 & 0      & 15 & 0.11 \\
\hline
\end{tabular}
\end{center}
\end{table}

\section{Discussion} \label{discussion}

We have presented a segmentation method based on penalized estimation and capable of processing multiple signals jointly. We have shown how this leads to  improvements in the analysis of normal samples (where segmentation can be applied to both total intensity and allelic proportion), tumor sample (where we are able to deal with contamination effectively), measurements from multiple platforms, and related individuals. Given that  copy number detection is such an active area of research, it is impossible to compare one method to all the others available. However, for each of the situations we analyzed, we tried to select approaches that represented the most successful state-of-the-art. In comparison to these, the algorithm we presented performs well: its accuracy is always comparable to that of the most effective competitor and its computation time often more contained. We believe that for its versatility and speed, GFL is particularly useful for initial screening. 

There are of course many aspects of CNV detection that we have not analyzed in this paper: from normalization and signal transformation to FDR control of detected CNVs. There are also a number of improvements to our approach that appear promising, but at this stage are left for further work: for example,  it is easy to modify algorithms so that the penalization parameters are location dependent to incorporate prior information on known copy number polymorphisms; more challenging is developing theory and method to select the values of these regularization parameters in a data-adaptive fashion. 

Finally, while our scientific motivation has been the study of copy number variations, the joint segmentation algorithm we present is not restricted to specific characteristics of these data types, and we expect it will be applied in other contexts.

\subsection*{Software implementation}

All the code used to run the analysis presented in this paper is available at the web-page of the authors. We have implemented the segmentation routine, which is our core contribution, in an R package ({\tt Piet}) to be submitted to R-forge (http://r-forge.r-project.org). To demonstrate a visualization of the CNV results on Chromosome 8 in the bipolar disorder study (see Section \ref{BP_study}),  we refer the interested audience to Supplementary Figure \ref{GFL_Chr8} in the supplementary material.

\section*{Acknowledgments}
The authors thank Nelson Freimer and all the collaborators in the Bipolar Endophenotype Mapping project for authorizing use of the genotype data. We also thank Susan Service and Joseph DeYoung for assistance in data management and interpretation, and Pierre Neuvial and Henrik Bengtsson for helpful discussion. C. S. gratefully acknowledges support from NIH/NIGMS GM053275 and MH075007 and K. L. from NIH/NIGMS GM053275.


\bibliographystyle{plain}
\bibliography{arxiv_v2}

\clearpage
\setcounter{page}{1}

\begin{center}

$\;\;$
\hrule
\hrule
$\;\;$

\vspace{1cm}
{\bf\large{Reconstructing DNA Copy Number by Joint Segmentation of Multiple Sequences}}

{\bf\large{\sc Supplementary Material}}
\vspace{0.5cm}

Zhongyang Zhang, Kenneth Lange, and Chiara Sabatti
\vspace{2cm}


\hrule
\hrule
\vspace{4cm}


March 2012

\end{center}

\newpage
\clearpage 
\renewcommand{\tablename}{Supplementary Table}
\setcounter{table}{0}

\renewcommand{\figurename}{Supplementary Figure}
\setcounter{figure}{0}

\renewcommand{\theequation}{S.\arabic{equation}}
\setcounter{equation}{0}

\subsection*{TDM algorithm}

The non-zero entries in $\mathbf{A}_i$ and $\mathbf{b}_i$ in the re-shaped surrogate function (\ref{quadratic_form}) are listed as follows:
\begin{eqnarray*}
a_i^{(m)}(1,1) & = & 1 + \frac{\lambda_{1,i}}{||\beta_{i1}^{(m)}||_{2,\epsilon}}
+ \frac{\lambda_{2,i}}{||\beta_{i2}^{(m)}-\beta_{i1}^{(m)}||_{2,\epsilon}} + \frac{\lambda_{3,i}^2}{||\boldsymbol{\lambda}_3 * (\boldsymbol{\beta}_{(2)}^{(m)} - \boldsymbol{\beta}_{(1)}^{(m)})||_{2,\epsilon}}; \\
a_i^{(m)}(j,j) & = & 1 + \frac{\lambda_{1,i}}{||\beta_{ij}^{(m)}||_{2,\epsilon}}
+ \frac{\lambda_{2,i}}{||\beta_{ij}^{(m)}-\beta_{i,j-1}^{(m)}||_{2,\epsilon}}
+ \frac{\lambda_{2,i}}{||\beta_{i,j+1}^{(m)}-\beta_{ij}^{(m)}||_{2,\epsilon}} \\
& & + \frac{\lambda_{3,i}^2}{||\boldsymbol{\lambda}_3 * (\boldsymbol{\beta}_{(j)}^{(m)} - \boldsymbol{\beta}_{(j-1)}^{(m)})||_{2,\epsilon}}
+ \frac{\lambda_{3,i}^2}{||\boldsymbol{\lambda}_3 * (\boldsymbol{\beta}_{(j+1)}^{(m)} - \boldsymbol{\beta}_{(j)}^{(m)})||_{2,\epsilon}},\\
& & \quad\quad\quad\quad\quad\quad\quad\quad\quad\quad\quad\quad\quad\quad \quad\quad\quad\quad j=2,\ldots,n-1;\\
a_i^{(m)}(n,n) & = & 1 + \frac{\lambda_{1,i}}{||\beta_{in}^{(m)}||_{2,\epsilon}}
+ \frac{\lambda_{2,i}}{||\beta_{in}^{(m)}-\beta_{i,n-1}^{(m)}||_{2,\epsilon}} + \frac{\lambda_{3,i}^2}{||\boldsymbol{\lambda}_3 * (\boldsymbol{\beta}_{(n)}^{(m)} - \boldsymbol{\beta}_{(n-1)}^{(m)})||_{2,\epsilon}}; \\
a_i^{(m)}(j,j-1) & = & -\frac{\lambda_{2,i}}{||\beta_{ij}^{(m)}-\beta_{i,j-1}^{(m)}||_{2,\epsilon}} - \frac{\lambda_{3,i}^2}{||\boldsymbol{\lambda}_3 * (\boldsymbol{\beta}_{(j)}^{(m)} - \boldsymbol{\beta}_{(j-1)}^{(m)})||_{2,\epsilon}}, \quad j=2,\ldots,n; \\
a_i^{(m)}(j,j+1) & = & -\frac{\lambda_{2,i}}{||\beta_{i,j+1}^{(m)}-\beta_{ij}^{(m)}||_{2,\epsilon}} - \frac{\lambda_{3,i}^2}{||\boldsymbol{\lambda}_3 * (\boldsymbol{\beta}_{(j+1)}^{(m)} - \boldsymbol{\beta}_{(j)}^{(m)})||_{2,\epsilon}}, \quad j=1,\ldots,n-1; \\
b_i^{(m)}(j) & =&  y_{ij},\quad j=1,\ldots,n.
\end{eqnarray*}

When staking measurements at different positions, the item $1$ in $a_i^{(m)}(j,j)$ is replaced by $\delta_{ij}$ and $b_i^{(m)}=y_{ij}$ is replaced by $b_i^{(m)}=\delta_{ij}y_{ij}$.

\subsection*{Bias estimation}
Let $x_{ij}$ be the data for sequence $i$ at locus $j$ after $\sigma_i$ of each sequence is normalized to $1$. With such normalization, the model (\ref{mmodel}) is reduced to a simpler form with global tuning parameters to each sequence for easier interpretation:
\begin{equation} \label{model_simple}
f(\boldsymbol{\beta}) = \frac{1}{2}\sum_{i=1}^M\sum_{j=1}^N (x_{ij}-\beta_{ij})^2 + \lambda_1 \sum_{i=1}^M  \sum_{j=1}^N |\beta_{ij}| + \lambda_2 \sum_{i=1}^M\sum_{j=2}^N |\beta_{ij} - \beta_{i,j-1}| +  \lambda_3 \sum_{j=2}^N \left[\sum_{i=1}^M (\beta_{ij}-\beta_{i,j-1})^2\right]^{\frac{1}{2}}.
\end{equation}
The solution to minimize $f(\boldsymbol{\beta})$ is unique for $f(\boldsymbol{\beta})$ is strictly convex. Denote the solution as $\hat{\boldsymbol{\beta}}=(\hat{\beta}_{ij})_{M\times N}$. Suppose sequence $i$ is partitioned into $\hat{K}_i$ consecutive segments $\{\hat{R}_1^{(i)},\ldots,\hat{R}_{\hat{K}_i}^{(i)}\}$, delimited with change points $\hat{\mathcal{J}_i}=\{\hat{j}_1^{(i)},\ldots, \hat{j}_{\hat{K}_i-1}^{(i)}\} \subset \{2,\ldots,N\}$ (left end of segment $2,\ldots,\hat{K}_i$). The fitted means of each segment is denoted as $\hat{\boldsymbol{\mu}}^{(i)}=(\hat{\mu}_1^{(i)},\ldots,\hat{\mu}_{\hat{K}_i}^{(i)})$, i.e., $\hat{\beta}_{ij}=\hat{\mu}_k^{(i)}$, if $j \in \hat{R}_k^{(i)}$. The length (number of SNPs) of each segment is $\hat{L}_k^{(i)}=|\hat{R}_k^{(i)}|,k=1,\ldots,\hat{K}_i$. Thus, the estimated mean vector for sequence $i$ can be written as
$$
\hat{\boldsymbol{\beta}}_i =  \sum_{k=1}^{\hat{K}_i} \hat{\mu}_k^{(i)}I_{\hat{R}_k^{(i)}}.
$$

$\hat{\boldsymbol{\beta}}$ is the optimal solution if and only if it satisfies the subgradient condition $\partial f(\hat{\boldsymbol{\beta}}) = 0$; that is,
\begin{equation} \label{subgrad_cond}
\hat{\beta}_{ij} = y_{ij} - \lambda_1 s_{ij}^{(1)} - \lambda_2 s_{ij}^{(2)} - \lambda_3 s_{ij}^{(3)} ,
\end{equation}
where $s_{ij}^{(1)}$, $s_{ij}^{(2)}$ and $s_{ij}^{(3)}$  are coordinates of subgradient corresponding to $\beta_{ij}$'s appearing in each of the three penalty terms. Both bias estimation and asymptotic analysis rely on the analytic form of subgradient. Now we discussed the bias induced by each penalty separately.

\subsubsection*{Bias induced by lasso penalty}
It is easy to verify that the subgradient for the lasso penalty  can be written as
$$
s_{ij}^{(1)} = \sign(\beta_{ij}),
$$
where, with a bit abuse of notation,
\begin{equation} \label{def_sign}
\sign(x) = \left\{\begin{array}{ll}
1, & \mbox{if } x>0,\\
-1, & \mbox{if } x<0,\\
z \in [-1,1], & \mbox{if } x=0.
\end{array}\right.
\end{equation}
Hence, the lasso penalty term merely plays as a soft-thresholding on the fitted values resulted from the model (\ref{model_simple}) with $\lambda_1=0$, denoted as $\hat{\beta}_{ij}(0,\lambda_2,\lambda_3)$; that is, for any $\lambda_1>0$,
$$
\hat{\beta}_{ij}(\lambda_1,\lambda_2,\lambda_3) = \sign\left[\hat{\beta}_{ij}(0,\lambda_2,\lambda_3)\right]\left[\hat{\beta}_{ij}(0,\lambda_2,\lambda_3) - \lambda_1\right]_{+},
$$
where $(x)_{+} = \max\{x,0\}$. This is also highlighted in Lamma A.1 of \cite{pco} for model (\ref{model_simple}) with $\lambda_3=0$.

\subsubsection*{Bias induced by fused-lasso penalty}

In model (\ref{model_simple}) with $\lambda_1=0$ and $\lambda_3=0$ (only fused-lasso penalty involved), Lemma 2.1 in \cite{FL_AOS} gives an insightful characterization of $\hat{\boldsymbol{\mu}}^{(i)}$:
$$
\hat{\mu}_k^{(i)} = \frac{1}{\hat{L}_k^{(i)}}\sum_{j \in \hat{R}_k^{(i)}}x_{ij} + \hat{c}_k^{(i)},\quad k=1,\ldots,\hat{K}_i,
$$
where
$$
\hat{c}_1^{(i)} =
\left\{\begin{array}{ll}
-\frac{\lambda_2}{\hat{L}_1^{(i)}}, & \mbox{if } \hat{\mu}_2^{(i)} - \hat{\mu}_1^{(i)}>0,\\
\frac{\lambda_2}{\hat{L}_1^{(i)}}, & \mbox{if } \hat{\mu}_2^{(i)} - \hat{\mu}_1^{(i)}<0,
\end{array}\right.
$$
$$
\hat{c}_{\hat{K}_i}^{(i)} =
\left\{\begin{array}{ll}
\frac{\lambda_2}{\hat{L}_{\hat{K}_i}^{(i)}}, & \mbox{if } \hat{\mu}_{\hat{K}_i}^{(i)} - \hat{\mu}_{\hat{K}_i-1}^{(i)}>0,\\
-\frac{\lambda_2}{\hat{L}_{\hat{K}_i}^{(i)}}, & \mbox{if } \hat{\mu}_{\hat{K}_i}^{(i)} - \hat{\mu}_{\hat{K}_i-1}^{(i)}<0,
\end{array}\right.
$$
and, for $k=2,\ldots,\hat{K}_i-1$,
$$
\hat{c}_k^{(i)} =
\left\{\begin{array}{ll}
\frac{2\lambda_2}{\hat{L}_k^{(i)}}, & \mbox{if } \hat{\mu}_k^{(i)} - \hat{\mu}_{k-1}^{(i)}<0, \hat{\mu}_{k+1}^{(i)} - \hat{\mu}_{k}^{(i)}>0, \\
-\frac{2\lambda_2}{\hat{L}_k^{(i)}}, & \mbox{if } \hat{\mu}_k^{(i)} - \hat{\mu}_{k-1}^{(i)}>0, \hat{\mu}_{k+1}^{(i)} - \hat{\mu}_{k}^{(i)}<0, \\
0, & \mbox{if } (\hat{\mu}_k^{(i)} - \hat{\mu}_{k-1}^{(i)})(\hat{\mu}_{k+1}^{(i)} - \hat{\mu}_{k}^{(i)})>0.
\end{array}\right.
$$
The result implies that the sample mean (as an unbiased estimate of true mean) of a local minimum/maximum segment (except it is located at either end) is shifted towards $0$ due to fused-lasso penalty. The bias is positively proportional to $\lambda_2$ and negatively proportional to the length of the segment. It is more important to notice that there exists no configuration where a local minimum/maximum segment has a jump size (relative to neighboring segments) less than the amount of bias. It means that a CNV with small jump size or small length could possibly be merged into neighboring segments, if $\lambda_2$ is set too large. 

\subsubsection*{Bias induced by group-fused-lasso penalty}

The subgradient for group-fused-lasso penalty is given in the following Proposition 1.

\textbf{Proposition 1}: The $\beta_{ij}$'s involved in group-fused-lasso penalty have subgradient given by
\begin{equation} \label{def_s3}
s_{ij}^{(3)} = \left\{\begin{array}{ll}
-e_{i2}, & \mbox{if } j=1,\\
e_{ij} - e_{i,j+1}, & \mbox{if } 1<j<N,\\
e_{iN}, & \mbox{if } j=N,
\end{array}\right.
\end{equation}
for $i=1,\ldots,M$, where $\mathbf{e}_j=(e_{1j},\ldots,e_{Mj})^T$ for $j=2,\ldots,M$ are given by
\begin{equation} \label{def_e}
\mathbf{e}_j = \left\{\begin{array}{ll}
\left(\frac{\beta_{1j}-\beta_{1,j-1}}{||\boldsymbol{\beta}_{(j)}-\boldsymbol{\beta}_{(j-1)}||_{\ell_2}},\ldots,
\frac{\beta_{Mj}-\beta_{M,j-1}}{||\boldsymbol{\beta}_{(j)}-\boldsymbol{\beta}_{(j-1)}||_{\ell_2}}\right)^T, & \mbox{if } ||\boldsymbol{\beta}_{(j)}-\boldsymbol{\beta}_{(j-1)}||_{\ell_2}>0,\\
\mbox{any } (e_{1j},\ldots,e_{Mj})^T \mbox{ s.t. } ||\mathbf{e}_j||_{\ell_2}\leq1 , & \mbox{if } ||\boldsymbol{\beta}_{(j)}-\boldsymbol{\beta}_{(j-1)}||_{\ell_2}=0.\\
\end{array}\right.
\end{equation}\\

\emph{Proof}: The proof follows a similar technique used in the proof of Lamma A.1 in \cite{FL_AOS}. Let $\mathbf{T}=[-\mathbf{I}_M,\mathbf{I}_M]$, where $\mathbf{I}_M$ is $M \times M$ identity matrix. Then, for any $2 \leq j \leq N$, 
$$
h(\boldsymbol{\beta}_{(j-1)},\boldsymbol{\beta}_{(j)})  \triangleq ||\boldsymbol{\beta}_{(j)}-\boldsymbol{\beta}_{(j-1)}||_{\ell_2} 
= ||\mathbf{T}[\boldsymbol{\beta}_{(j-1)}^T,\boldsymbol{\beta}_{(j)}^T]^T||_{\ell_2}.
$$
For the $j$ such that $||\boldsymbol{\beta}_{(j)}-\boldsymbol{\beta}_{(j-1)}||_{\ell_2} >0$, the sub-gradient is reduced to regular gradient, and thus can be derived in a usual way. We now focus on the $j$ such that $||\boldsymbol{\beta}_{(j)}-\boldsymbol{\beta}_{(j-1)}||_{\ell_2}=0$, i.e., the subgradient of $\beta_{ij}$ at $0$. By Cauchy-Schwartz inequality, we have
$$
\begin{array}{rcl}
h(\boldsymbol{\beta}_{(j-1)},\boldsymbol{\beta}_{(j)})  &\geq& ||\mathbf{T}[\boldsymbol{\beta}_{(j-1)}^T,\boldsymbol{\beta}_{(j)}^T]^T||_{\ell_2}||\mathbf{e}_j||_{\ell_2} \\
&\geq& <\mathbf{T}[\boldsymbol{\beta}_{(j-1)}^T,\boldsymbol{\beta}_{(j)}^T]^T,\mathbf{e}_j> \\
&=& h(\mathbf{0}) + <[\boldsymbol{\beta}_{(j-1)}^T,\boldsymbol{\beta}_{(j)}^T]^T-\mathbf{0},\mathbf{T}^T\mathbf{e}_j>
\end{array}
$$
where $\mathbf{e}_j$ is any vector such that $||\mathbf{e}_j||_{\ell_2} \leq 1$. It follows by the definition of subgradient that $\mathbf{T}^T\mathbf{e}_j=[-\mathbf{e}_j^T,\mathbf{e}_j^T]^T$ is the subgradient for $[\boldsymbol{\beta}_{(j-1)}^T,\boldsymbol{\beta}_{(j)}^T]^T$. $\square$\\

The bias induced by the group-fused-lasso penalty can be derived from the analytic form of subgradient accordingly and is given in the following Proposition 2.

\textbf{Proposition 2}: In model (\ref{model_simple}) with $\lambda_1=0$ and $\lambda_2=0$, the fitted means of segments for sequence $i$ can be expressed as
$$
\hat{\mu}_k^{(i)} = \frac{1}{\hat{L}_k}\sum_{j \in \hat{R}_k^{(i)}}x_{ij} + \hat{c}_k^{(i)},\quad k=1,\ldots,\hat{K}_i,
$$
where
$$
\hat{c}_k^{(i)} =
\left\{\begin{array}{ll}
\frac{\lambda_3}{\hat{L}_1^{(i)}} \cdot  r_i(\hat{j}_1^{(i)}), & \mbox{if } k=1, \\
-\frac{\lambda_3}{\hat{L}_k^{(i)}} \cdot \left[r_i(\hat{j}_{k-1}^{(i)}) - r_i(\hat{j}_k^{(i)})\right], & \mbox{if } 2\leq k \leq \hat{K}_i-1, \\
-\frac{\lambda_3}{\hat{L}_{\hat{K}_i}^{(i)}} \cdot  r_i(\hat{j}_{\hat{K}_i-1}^{(i)}), & \mbox{if } k=\hat{K}_i,
\end{array}\right.
$$
and
$$
r_i(j) \triangleq \frac{\hat{\beta}_{ij}-\hat{\beta}_{i,j-1}}{||\hat{\boldsymbol{\beta}}_{(j)}-\hat{\boldsymbol{\beta}}_{(j-1)}||_{\ell_2}}.
$$\\

\emph{Proof}: The proof follows a similar technique used in the proof of Lemma 2.1 in \cite{FL_AOS}.  Following the subgradient condition (\ref{subgrad_cond}) in case $\lambda_1=0$ and $\lambda_2=0$, we have
$$
\hat{\mu}_k^{(i)} = \frac{1}{\hat{L}_k}\sum_{j \in \hat{R}_k^{(i)}}\hat{\beta}_{ij} = \frac{1}{\hat{L}_k}\sum_{j \in \hat{R}_k^{(i)}}x_{ij} - \frac{\lambda_3}{\hat{L}_k}\sum_{j \in \hat{R}_k^{(i)}}s_{ij}^{(3)}.
$$
By Proposition 1 and simple algebra, we have
$$
\sum_{j \in \hat{R}_k^{(i)}}s_{ij}^{(3)} =
\left\{\begin{array}{ll}
-e_{i,\hat{j}_1^{(i)}}, & \mbox{if } k=1, \\
e_{i,\hat{j}_{k-1}^{(i)}} - e_{i,\hat{j}_k^{(i)}}, & \mbox{if } 2\leq k \leq \hat{K}_i-1, \\
e_{i,\hat{j}_{\hat{K}_i-1}^{(i)}}, & \mbox{if } k=\hat{K}_i.
\end{array}\right.
$$
Note that at jump points, subgradient has explicit form as shown in Proposition 1. It follows that $e_{i,\hat{j}_k^{(i)}}=r_i(\hat{j}_k^{(i)})$, for $k=1,\ldots,\hat{K}_i-1$, where $r_i(\cdot)$ is defined in Proposition 2. $\square$\\

Some interesting implications follow immediately. For sequence $i$, consider one of its fitted segment $k$ with end points $[\hat{j}_{k-1}^{(i)},\hat{j}_k^{(i)}-1]$. If no other sequences share change points at these two ends, then the bias term $\hat{c}_k^{(i)}$ reduces to what it appears in model (\ref{model_simple}) with fused-lasso term only ($\lambda_1=0$ and $\lambda_3=0$). If $m$ out of $M$ sequences share change points at these two ends and also assume the jump size at these two locations for all the $m$ sequences are roughly the same, then the absolute value of the bias term can be approximately written as $\frac{2\lambda_3}{\hat{L}_k^{(i)}}\cdot\frac{1}{\sqrt{m}}$. It means that if more than one sequences share change points at the same coordinate, then they can benefit from each other to reduce their individual bias, relative to the bias induced by fused-lasso penalty specific to each individual sequence.

\subsection*{Asymptotic behavior}

Now we try to give a justification of the order of the magnitude of $\lambda_2$ and $\lambda_3$ in compatible with their large sample behavior, say, as $N \to \infty$. When the number of sequences $M$ in segmentation task is relatively large, extra caution is needed for $\lambda_3$. Again, we discuss asymptotic behavior of the solution influenced by fused-lasso and group-fused-lasso separately for easier exhibition. 

\subsubsection*{Asymptotic behavior for fused-lasso penalty}

In fused-lasso model ($\lambda_1=0$ and $\lambda_3=0$), the justification is directly inspired by the proof of Theorem 2.3 in \cite{FL_AOS}. Denote the event
$$
\mathcal{E}_i =\{\hat{\mathcal{J}}_i=\mathcal{J}_i\} \cap \{\sign(\hat{\beta}_{ij}-\hat{\beta}_{i,j-1})=\sign(\beta_{ij}-\beta_{i,j-1}), \forall j \in \mathcal{J}_i\},
$$
for $i=1,\ldots,M$ respectively. This event means that all jump points and the direction of jumps are correctly identified for each sequence $i$. A necessary condition required for $\lambda_2$ is summarized in Proposition 3.

{\bf Proposition 3}: It is required that $\lambda_2 = O(\sqrt{\log N})$ to ensure $\lim_{N \to \infty}\mathbb{P}(\mathcal{E}_i)=1$ for $i=1,\ldots,M$, at the linear rate.

This asymptotic behavior follows directly the proof of Theorem 2.3 in \cite{FL_AOS}. We have some quick remarks:
\begin{enumerate}
\item[1)]
If the signal of each sequence is not normalized, then $\lambda_{2,i}=c_2\sigma_i\sqrt{\log N}$, specific to sequence $i$.
\item[2)]
In order to ascertain a CNV segment with length $L$ and jump size $\delta$, the bias needs to satisfy $\frac{2\lambda_{2,i}}{L}=\frac{2c_2\sigma_i\sqrt{\log N}}{L}<\delta$, i.e.,  $c_2 < \frac{1}{2\sqrt{\log N}} \cdot \frac{\delta}{\sigma_i}L$. Here, $\frac{\delta}{\sigma_i}$ can be interpreted as signal-to-noise ratio (SNR). For a specific platform, one may get a sense of the magnitude of SNR and $L$ from prior knowledge. In practice, it is desired to take as large value of $c_2$ as possible to ensure the sparsity of the segmentation, but not too large in order to compensate for the constraint of signal strength ($\frac{\delta}{\sigma_i}L$). Based on our experiences of analysis of Illumina data \cite{FL_DPI}, the results are not sensitive to the choice of $c_2$, provided that it falls into a reasonable range.
\end{enumerate}

\subsubsection*{Asymptotic behavior for group-fused-lasso penalty}

In group-fused-lasso model ($\lambda_1=0$ and $\lambda_2=0$), we have similar requirement of $\lambda_3$ as for $\lambda_2$, which is given in Proposition 4.

\textbf{Proposition 4}: It is required that $\lambda_3 = O(\sqrt{M}\sqrt{\log N})$ to ensure $\lim_{N \to \infty}\mathbb{P}(\cap_{i=1}^M\mathcal{E}_i)=1$, at the linear rate.\\

\emph{Proof}: For simplicity, we prove under the condition that $\epsilon_{ij}$ are i.i.d. $\mathcal{N}(0,1)$ (after $\sigma_i$ is normalized to 1), while this condition can be relaxed \cite{FL_AOS}. We also follow the same technique used in the proof of Theorem 2.3 in \cite{FL_AOS}. Let $d_{ij}=\beta_{ij}-\beta_{i,j-1}$, $\hat{d}_{ij}=\hat{\beta}_{ij}-\hat{\beta}_{i,j-1}$, and $d_{ij}^{\epsilon}=\epsilon_{ij}-\epsilon_{i,j-1}$. Also denote $\mathbf{d}_{j}^{\epsilon} = (d_{1j}^{\epsilon},\ldots,d_{Mj}^{\epsilon})^T$ and $\mathcal{J} = \cup_{i=1}^M \mathcal{J}_i$.  By the subgradient condition (\ref{subgrad_cond}),  for each $i$, $\mathcal{E}_i$ holds if and only if
\begin{equation} \label{GFL_non_change_point_cond}
d_{ij}^{\epsilon} = \lambda_3(2e_{ij} - e_{i,j-1} - e_{i,j+1}),\quad \mbox{for } j \in \mathcal{J}_i^c,
\end{equation}
and
\begin{equation} \label{GFL_change_point_cond}
|\hat{d}_{ij}| > 0,\quad \mbox{for } j \in  \mathcal{J}_i.
\end{equation}
Condition (\ref{GFL_change_point_cond}) has direct relevance to the bias issue, as discussed above. Now we focus on condition (\ref{GFL_non_change_point_cond}), which implies that
$$
\max_{j \in \mathcal{J}^c} ||\mathbf{d}_{j}^{\epsilon}||_{\ell_2} = \max_{j \in \mathcal{J}^c} \lambda_3||2\mathbf{e}_{j} - \mathbf{e}_{j-1} - \mathbf{e}_{j+1}||_{\ell_2} < 4\lambda_3.
$$
It is left to show that $\mathbb{P}(\max_{j \in \mathcal{J}^c} ||\mathbf{d}_j^{\epsilon}||_{\ell_2} \geq4\lambda_3) = \mathbb{P}(\max_{j \in \mathcal{J}^c} ||\mathbf{d}_j^{\epsilon}/\sqrt{2}||_{\ell_2}^2 \geq 8\lambda_3^2)\to 0$ as $N \to \infty$ for $i=1,\ldots,M$. Note that for each $j$, $d_{1j}^{\epsilon},\ldots,d_{Mj}$ are i.i.d. $\mathcal{N}(0,2)$, so $||\mathbf{d}_{j}^{\epsilon}/\sqrt{2}||_{\ell_2}^2 \sim \chi_M^2$. Then we have
\begin{eqnarray*}
&&\mathbb{P}(\max_{j \in \mathcal{J}^c} ||\mathbf{d}_j^{\epsilon}/\sqrt{2}||_{\ell_2}^2 \geq 8\lambda_3^2) \\
&=& \mathbb{P}(\cup_{j \in \mathcal{J}^c} ||\mathbf{d}_j^{\epsilon}/\sqrt{2}||_{\ell_2}^2 \geq 8\lambda_3^2) \\
&\leq& \sum_{j \in \mathcal{J}^c} \mathbb{P}(||\mathbf{d}_j^{\epsilon}/\sqrt{2}||_{\ell_2}^2 \geq 8\lambda_3^2) \\
&=& |\mathcal{J}^c|  \mathbb{P}(||\mathbf{d}_j^{\epsilon}/\sqrt{2}||_{\ell_2}^2 \geq 8\lambda_3^2) \\
&\leq& \exp\left[-\frac{1}{2}(8\lambda_3^2 - M) + \log |\mathcal{J}^c| - \frac{M}{2}\log\frac{M}{8\lambda_3^2}\right].
\end{eqnarray*}
Here the first inequality is due to union bound and the second inequality is due to Chernoff's bound for $\chi_M^2$ distribution. Under the assumption on sparsity of the change points, we have $|\mathcal{J}^c|=O(N)$ for fixed $M$.  In our settings, $M$ is fixed (which may rise up to thousands) while $N \to \infty$, yet in practice, $M$ is not negligible with respect to $\sqrt{\log N}$. For example, $\sqrt{{\log(10^6)}} \approx 3.72$, and it is not uncommon to have more than $4$ sequences for joint segmentation. Therefore, it is necessary to have $\lambda_3 = O(\sqrt{M}\sqrt{\log N})$. $\square$\\

We also have some remarks on how to determine $\lambda_3$:
\begin{enumerate}
\item[1)]
If the signal of each sequence is not normalized, then $\lambda_{3,i}=c_3\sigma_i\sqrt{pM}\sqrt{\log N}$. The choice of $p$ is decided case by case and discussed in the main text.
\item[2)]
Following the above discussion about bias induced by group-fused-lasso penalty,  if $m$ out of $M$ sequences carry CNVs with exactly the same boundary,  the bias can be approximately written as $\frac{2c_3\sigma_i\sqrt{\log N}}{\hat{L}_k^{(i)}}\cdot\frac{\sqrt{pM}}{\sqrt{m}}$. On one hand, if $p$ is over estimated so that $pM$ is much larger than $m$, the model would be over penalized and introduce more bias than that is attributed to individual fused-lasso penalty, and thus does not benefit from joint analysis; On the other hand, if $pM$ is set too small, we have insufficient control on the sparsity of each sequence, so that it has to be compensated by the fused-lasso penalty. This is the reason why we need to incooperate $\rho(p)$ to re-weight the relative influence of the two penalties.
\end{enumerate}

\subsection*{Details in calling procedure}

We specify the likelihood functions of LRR and BAF signals in the log-likelihood ratio (\ref{lr}) as follows. For BAF signal, the likelihood is usually modeled for different copy number states as a mixture of densities surrounding a few possible BAF values corresponding to different genotypes \cite{QuantiSNP, PennCNV}. When population frequencies for allele A and B, $p_A$ and $p_B$, are available or can be estimated from data, we have
$$
L_{\mbox{\scriptsize BAF}}(x; c) = \sum_{s=0}^c {c \choose s}p_A^{c-s}p_B^s \phi_s(x;\mu_s,\sigma_s^2),\quad \mbox{for } c=0,1,2,3,4,
$$
where $\phi_s(\cdot;\mu_s,\sigma_s^2)$ is normal density for state $s$. The details in model and parameter specification are listed in Supplementary Table \ref{BAF_par}. 

In case where population frequencies $p_A$ and $p_B$ are not available, we might use an alternative likelihood function for BAF \cite{FL_DPI}, defined by
$$
L_{\mbox{\scriptsize BAF}}(x; c) = \max_{s \in \{0,\ldots, c\}} \phi_s(x;\mu_s,\sigma_s^2),\quad \mbox{for } c=0,1,2,3,4,
$$
where all parameters are defined in the same way (see Supplementary Table \ref{BAF_par}).\\

\begin{table}[htb]
\begin{center}
\begin{tabular}{cccccc}
\hline
$c$ & s & Genotype & $\phi_s(\cdot)$  & $\mu_s$ & $\sigma_s$ \\
\hline
0      & 0 & Null            &  normal               & $1/2$      & $10\hat{\sigma}_x$ \\
\hline
1      & 0, 1 & A, B                 &  half normal       & $0$, $1$         & $\hat{\sigma}_x$ \\
\hline
2      & 0, 2 & AA, BB               &  half normal       & $0$, $1$         & $\hat{\sigma}_x$ \\
        & 1 & AB               &  normal               & $1/2$      & $\hat{\sigma}_x$ \\
\hline
3      & 0, 3 & AAA, BBB            &  half normal       & $0$, $1$         & $\hat{\sigma}_x$ \\
        & 1, 2 & AAB, ABB            &  normal               & $1/3$, $2/3$      & $\hat{\sigma}_x$ \\
\hline
4      & 0, 4 & AAAA, BBBB            &  half normal       & $0$, $1$          & $\hat{\sigma}_x$ \\
        & 1, 2, 3 & AAAB, AABB, ABBB            &  normal               & $1/4$, $1/2$, $3/4$       & $\hat{\sigma}_x$ \\
\hline
\end{tabular}
\end{center}
\caption{Model and parameter specification in BAF signal for each copy number state. $\hat{\sigma}_x$ is empirically estimated from BAF values in $(0.4,0.6)$ for each individual.}
\label{BAF_par}
\end{table}

For LRR signal, the likelihood function is simply defined by normal density:
$$
L_{\mbox{\scriptsize LRR}}(y; c) = \phi(y;\mu_c,\sigma_c^2).
$$ 
For $c=0,1,3,4$, $\mu_c$ and $\sigma_c^2$ are estimated based on the data $\mathbf{y}_R$ in segment $R$ being considered, while $\mu_2$ and $\sigma_2^2$ are estimated from the data of the whole chromosome on which segment $R$ locates or, locally, from the data of a few hundred markers flanking the segment. 

\subsection*{Additional Results}

\begin{table}[htb]
\begin{center}
\label{region10}
\begin{tabular}{rlrrrrr}
\hline
Region &  Aberration Type & Chr & bp Start        & bp End      & \#SNP   & \#hetSNP \\
\hline
  1 &               CN-LOH & 5     &                    1 &   47700000 &   9397 &      2756 \\
  2 &                      Loss & 5     & 111789971 & 112521346 &   156 &          79 \\
  3 &                       Gain & 8     &                    1 &   45200000  &  12564 &     3830 \\
  4 &                      Gain  & 8    & 128432670 &  129207869 &   218 &          91 \\
  5 &                      Loss  & 9    &                    1 &    50600000 &   11201 &     3889 \\
  6 &                      Loss  & 10  &  84504379  &    94825178  &   1988 &       648 \\
  7 &                      Gain & 12  &                    1 & 132449811  &   27131 &     8818 \\
  8 &                      Loss & 13   &  31766569  &   31892852  &    37 &         10 \\
  9 &                CN-LOH & 17   &    7431864  &    11747138  &   1150  &       308 \\
10 &                CN-LOH & 17   &  22300000  &   78774742   &   9713 &    3205 \\
\hline
\multicolumn{6}{l}{Total number of modified heterozygous SNPs} & 23634 \\
\multicolumn{6}{l}{Total number of heterozygous SNPs on autosome} & 176207 \\
\multicolumn{6}{l}{Total number of SNPs on autosome} & 547359 \\
\hline
\end{tabular}
\end{center}
\caption{Regions of allelic imbalance imposed to the HapMap sample NA06991 \cite{Staaf08}.}
\label{CNVdescription}
\end{table}

\begin{table}[htb]
\begin{center}
\begin{tabular}{rl}
\hline
Method & Time per sample in sec. (mean (std dev)) \\
\hline
GFL  & 21.97 (1.31) \\
BAFsegmentation  & 41.73 (-) \\
PSCN & 1154.18 (74.73) \\
\hline
\end{tabular}
\caption{Speed comparison of three methods: GFL, BAFsegmentation and PSCN.} \label{tumor_time}
\end{center}
\end{table}

\begin{table}[htb]
\begin{center}
{\footnotesize
\begin{tabular}{l|lll|cccccc}
\hline
Sample & Gender & Ancestry & Resource & Type  &  $<$10k & 10$-$50k & 50$-$100k & $>$100k  & Total \\
\hline
  & & &  & loss & 12 & 25 & 3 & 7 & 47 \\
NA15510 & Female & European& PDR & gain & 0   & 0   &  1 & 4 & 5   \\
 & & &                  & total & 12 & 25 &  4 & 11 & 52 \\
\hline
 & & &   & loss & 10 & 22  & 4 & 4 & 40 \\
NA18517 & Female & YRI & HapMap & gain & 1   & 3    & 1 & 8  & 13 \\
 & & &                   & total & 11 & 25  & 5 & 12 & 53 \\
\hline
 & & &   & loss & 13 & 16 & 4 & 5 & 38 \\
NA18576 & Female & CHB & HapMap & gain & 0   & 2   & 2  & 4 & 8   \\
 & & &                  & total  & 13 & 18 & 6 & 9 & 46 \\
\hline
 & & &   & loss & 8 & 16 & 1 & 4 & 29 \\
NA18980 & Female & JPT & HapMap  & gain & 0 & 0   & 1 & 3 & 4    \\
 & & &                  & total & 8 & 16 & 2  & 7 & 33 \\
\hline
\end{tabular}
}
\end{center}\caption{Sample information and reference CNV regions summarized for each sample by their types and sizes. The ancestry of NA15510 was not recorded but inferred in \cite{NA15510_anc}. Abbreviation: PDR = Polymorphism Discovery Resource.}
\label{sample_info}
\end{table}

\begin{table}[htb]
\begin{center}
{\footnotesize
\begin{tabular}{c|cc|cc|cc|cc|cc|c}
\hline
                                        & & &  \multicolumn{2}{c|}{NA15510} &  \multicolumn{2}{c|}{NA18517} & \multicolumn{2}{c|}{NA18576} &  \multicolumn{2}{c|}{NA18980} & \\
\cline{2-9}
\hline
Analysis                            & $\rho$ & $M$ &   \# Det. & \# Ovlp.        &  \# Det. & \# Ovlp.          &  \# Det. & \# Ovlp         &  \# Det. & \# Ovlp  & Time (min.) \\
\hline
\hline
\multicolumn{10}{l}{Analysis A: GFL done on averaged signal for each platform} \\
\hline
O1Q     & 1 & 1 & 92      & 34                     &  73   & 22                                       &  71  & 21                                      &  69  & 20  &  0.3\\
O2Q      & 1 & 1 & 114    & 22                     &  92   & 24                                       &  111  & 15                                      &  95  & 11 &  0.9 \\
Union   & -  & - & 170    & 38                     & 144  & 34                                       &  160  & 25                                      &  145  & 22 & 1.2\\
\hline
\hline
\multicolumn{10}{l}{Analysis B: GFL done on averaged signal of both platforms jointly} \\
\hline
     & 0 & 2 & 128    & 40                     & 108  & 33                                        &  96  & 21                                      &  104 & 23 & 4.2 \\
\hline
\hline
\multicolumn{10}{l}{Analysis C: GFL done on three replicates separately for each platform} \\
\hline
O1Q        & 1 & 1 &  66      & 31                     &  65   & 22                                       &  43  & 19                                      &  48  & 15 & 0.9 \\
O2Q        & 1 & 1 &  68      & 23                     &  65   & 22                                       &  65  & 12                                      &  59  & 13 & 2.8\\
Union  & -  & -  & 102    & 36                     & 109  & 33                                       &  93  & 25                                      &  91  & 20 & 3.7 \\
\hline
\hline
\multicolumn{10}{l}{Analysis D: GFL done on three replicates jointly for each platform} \\
\hline
O1Q           & 0 & 3  &  64      & 32                     &  66    & 22                                       &  54  & 21                                      &  53 & 18  & 1.1\\
O2Q           & 0 & 3  &  75       & 22                     &  70    & 24                                       &  65  & 11                                      &  49 & 12  & 3.1\\
Union          & -  & -   &  106    & 36                     &  115 & 33                                        &  96  & 22                                      &  83 & 21  & 4.2\\
\hline
\hline
\multicolumn{10}{l}{Analysis E: GFL done on three replicates of both platforms jointly} \\
\hline
         & 0 & 6  &  80      & 38                     &  82   & 32                                        &  69  & 25                                      &  56 & 15  & 8.5 \\
\hline
\hline
\multicolumn{10}{l}{MPCBS: Segmentation done on three replicates of both platforms jointly} \\
\hline
                   & - & -  &  98      & 34                             &  88   & 28                                        &  59  & 18                                          & 68   & 21   & 313.9    \\
\hline
\end{tabular}}
\caption{Number of CNVs detected (Det.) and overlapping (Ovlp.) with reference results as well as average computation time for four samples under different analyses. Tuning parameters used in segmentation: $c_1=0.1$, $c_2=2$, $c_3=2$ and $p=1$; $\rho$ and $M$ are specified for each analysis. Analysis A, C and E correspond to Analysis 1, 2 and 3 respectively in Table \ref{multipltfm_res} of main text.}
\label{multipltfm_data}
\end{center}
\end{table}

\begin{figure}[htb]
\centering
\subfigure[Individual analysis]{
\includegraphics[width=0.45\textwidth,page=1]{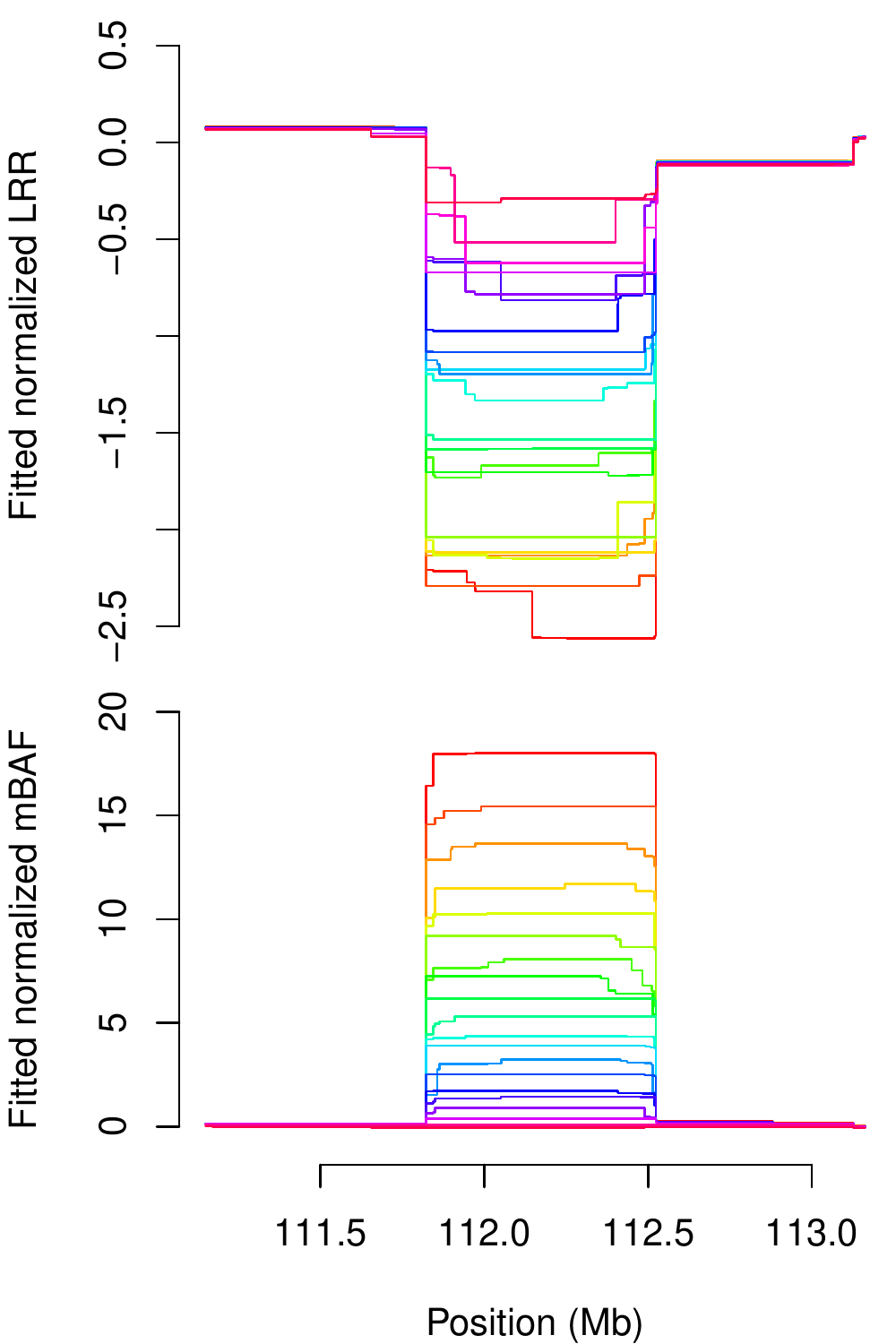}
}
\subfigure[Joint analysis]{
\includegraphics[width=0.45\textwidth,page=2]{./indv_joint.pdf}
}
\\
\subfigure{
\includegraphics[width=0.9\textwidth]{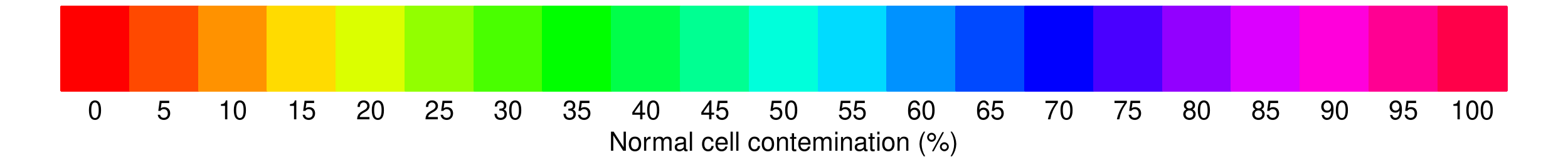}
}
\caption{Comparison of fitted profiles between analysis for each tumor sample with different normal cell contamination levels and joint analysis for all $21$ tumor samples. Shown is a hemizygous loss on Chromosome 5q22. In each of the subplots, the upper panel shows the fitted profiles on LRR for each sample distinctly marked by a spectrum of colors , while the lower panel shows their corresponding fitted profiles on mBAF. Shown are data points for heterozygous makers. (a) Individual analysis; (b) Joint analysis.}
\label{tumor_joint}
\end{figure}

\begin{figure}[htb]
\centering
\includegraphics[width=1\textwidth]{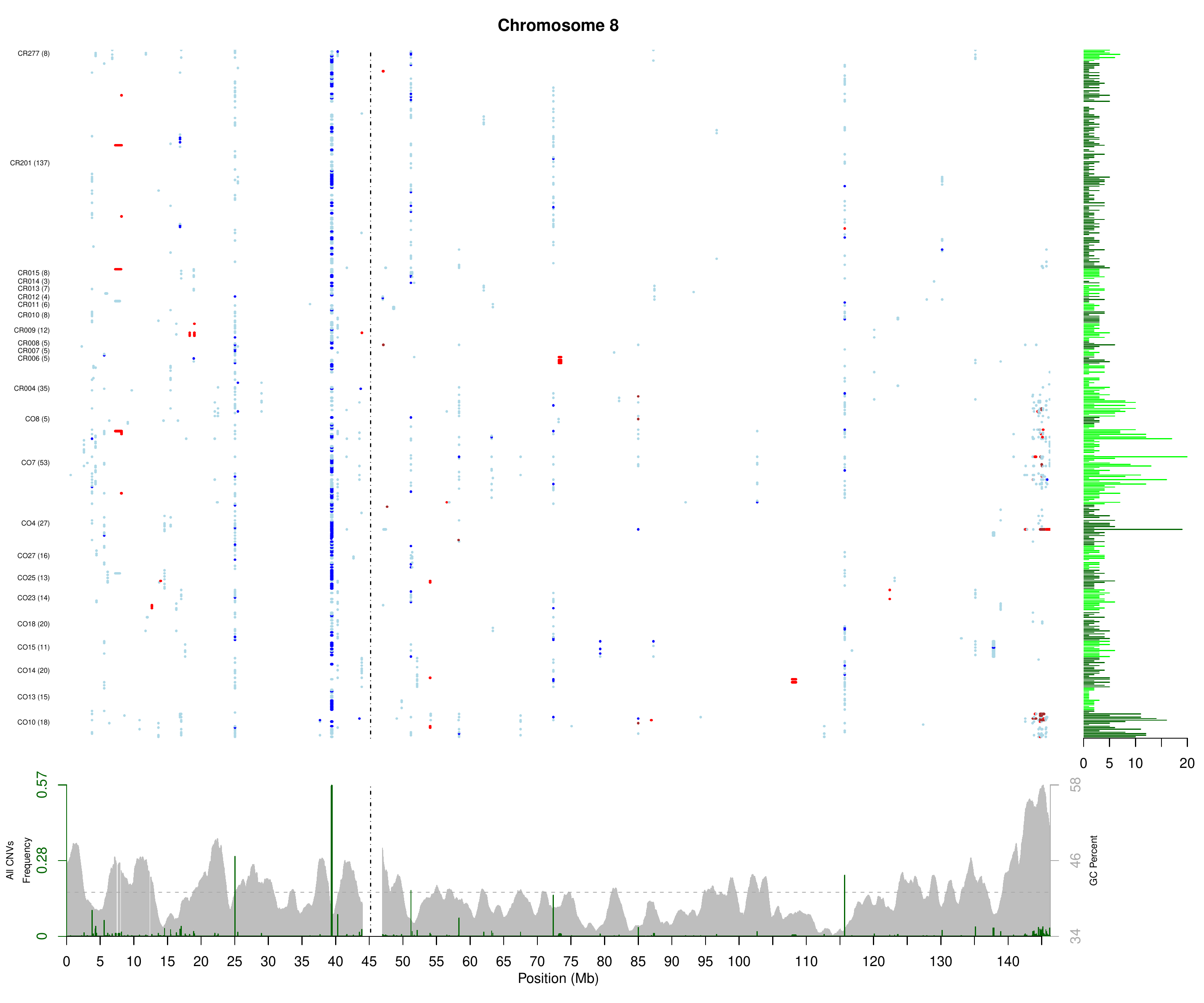}
\caption{Visualization of pedigree-wise CNV analysis results of Chromosome 8 data in bipolar disorder study. In the main body of the plot, CNVs estimated for each individual are marked by small segments with color code: CN=0 in blue, CN=1 in light blue, CN=3 in red and CN=4 in brown. Each subject is a row, each SNP a column. Subjects belonging to the same pedigree are stacked together. The pedigree names are indicated on the left hand side with the number of pedigree members included in parentheses. On the right hand side, the barplot represents the number of CNV detected per subject. Two shades of green are switched alternately to indicate the pedigree to which the subject belongs. At the bottom, the gray histogram shows the GC content along the chromosome; coordinated with the representation of CNVs in the main body,  the green histogram counts the frequency of CNV among the subjects represented. Vertical dotted line marks the centromere.}
\label{GFL_Chr8}
\end{figure}

\end{document}